\documentclass[apjl]{emulateapj}

\usepackage{epsfig}
\usepackage{verbatim}

\begin{document}

\slugcomment{Accepted for publication in ApJ, 1/24/06}
\shortauthors{TREMBLAY ET AL.}
\shorttitle{THE WARPED DISK OF 3C 449}

\title{The Warped Nuclear Disk of Radio Galaxy 3C 449}
 
\author{Grant R.~Tremblay\altaffilmark{1}}
\author{Alice C.~Quillen\altaffilmark{1}}
\author{David J.~E.~Floyd\altaffilmark{2}}
\author{Jacob Noel-Storr\altaffilmark{3}}
\author{Stefi A.~Baum\altaffilmark{4}}
\author{David Axon\altaffilmark{5}}
\author{Christopher P.~O'Dea\altaffilmark{5}}
\author{Marco Chiaberge\altaffilmark{2,6}}
\author{F.~Duccio Macchetto\altaffilmark{2}}
\author{William B.~Sparks\altaffilmark{2}}
\author{George K.~Miley\altaffilmark{7}}
\author{Alessandro Capetti\altaffilmark{8}}
\author{Juan P.~Madrid\altaffilmark{2}}
\author{Eric Perlman\altaffilmark{9}}

\altaffiltext{1}{Department of Physics and Astronomy,
University of Rochester, Rochester, NY 14627;
grant@pas.rochester.edu; aquillen@pas.rochester.edu}

\altaffiltext{2}{Space Telescope Science Institute, 3700 San Martin Dr.,
Baltimore, MD 21218; floyd@stsci.edu}

\altaffiltext{3}{Steward Observatory, University of Arizona, 
Tucson, AZ 85721}

\altaffiltext{4}{Center for Imaging Science, Rochester Institute of Technology, 
54 Lomb Memorial Drive, Rochester, NY 14627}

\altaffiltext{5}{Department of Physics, Rochester Institute of Technology,
54 Lomb Memorial Drive, Rochester, NY 14627}

\altaffiltext{6}{On leave from INAF---Istituto di Radioastronomia, Via P.~Gobetti 101, 
Bologna I-40129, Italy}

\altaffiltext{7}{Leiden Observatory, P.O.~Box 9513, NL-2300 RA Leiden, 
The Netherlands}

\altaffiltext{8}{INAF---Osservatorio Astronomico di Torino, 
Strada Osservatorio 20, 10025 Pino Torinese, Italy}

\altaffiltext{9}{Joint Center for Astrophysics, Department of Physics, 
University of Maryland, Baltimore County, 1000 Hilltop Circle, Baltimore, 
MD 21250}

\begin{abstract}
Among radio galaxies containing nuclear dust disks,
the bipolar jet axis is generally observed to be 
perpendicular to the disk major axis.
The FR I radio source 3C 449 is an outlier to this statistical 
majority, as it possesses a nearly parallel jet/disk orientation on the sky. 
We examine the 600 pc dusty disk in this galaxy with images from the
{\it Hubble Space Telescope}.
We find that a 1.6 $\mu$m/0.7 $\mu$m  colormap of the disk exhibits a twist
in its isocolor contours (isochromes).
We model the colormap by integrating galactic starlight 
through an absorptive disk, and find that the anomalous twist in the 
isochromes can be reproduced in the model with a vertically thin,
warped disk.  
The model predicts that 
the disk is nearly perpendicular to
the jet axis within 100 pc of the nucleus.
We discuss physical mechanisms capable of causing such a warp.
We show that precessional models or a torque on the disk arising from
a possible binary black hole in the AGN causes precession 
on a timescale that is too long to account for the predicted disk morphology. 
However, we estimate that the pressure in the X-ray emitting interstellar medium
is large enough to perturb the disk, and argue that jet-driven anisotropy in 
the excited ISM may be the cause of the warp. 
In this way, the warped disk in 3C 449 may
be a new manifestation of feedback from an active galactic nucleus.
\end{abstract}


\section{Introduction}

Imaging campaigns have established that radio-loud elliptical galaxies 
can host gaseous disks in their central few hundred parsecs, 
thought to be feeding the massive $10^9 M_\odot$ black 
holes at their centers (e.g., \citealt{ferrarese96,jaffe, ferrarese99}).
Visible and radio images of normal and radio-loud elliptical galaxies
can be used to probe the relation between the 
orientation of the jets 
and the morphology of the gas and dust near the nucleus.  
Ground based studies first discovered a connection, 
finding that dusty disks are often nearly perpendicular 
to radio jets \citep{mollenhoff92,kotanyi79}.  
These studies supported the widely held expectation 
that there is a link between 
the angular momentum of accreting material and the orientation 
of the jet axis.  Subsequent
surveys of the observed angular difference on the sky, $\Psi_{JD}$, 
between bipolar jet and disk major
axes in radio galaxies found a statistically significant
peak in the distribution at $\Psi_{JD} \approx 90^\circ$
\citep{mollenhoff92,kotanyi79,martel00,vandokkum95,dekoff00,kleijn99}. 
Taking 3D jet orientation into account and correcting for beaming effects, 
\citet{sparks} also showed
that these disks are generally orthogonal to jet axes. 
The correlation could be stronger for
FR I type sources than FR II types \citep{dekoff00, fanaroff74},
and as of yet, no strong trend between the jet axis and the galactic
isophotal major axis is known to exist in low redshift objects
(e.g., \citealt{sansom87,kinney}). 

More recent examinations of 3D jet/disk orientations 
have found additional complexities in the distribution.
The statistical work by \citet{kleijn} found
that dust {\it lanes} are nearly  perpendicular to jets,
whereas dusty disks with ellipsoidal edges have a wide range of
intrinsic jet/disk orientation angles.
\citet{schmitt} showed that 
projection effects of radio and gaseous structures on the sky must be taken into 
account in any such study of jet/disk orientations. 
This work demonstrated that the statistical distribution of $\Psi_{JD}$ 
is consistent with a homogeneous three dimensional angular distribution that
is modified by including a cone of avoidance.
The cone corresponds to an absence of 
jet axes lying within $\sim 13-25^\circ$ 
of the plane of the disk.
This may suggest that jet/disk orthogonality  
is not a trend, but simply a manifestation of the cone of avoidance in the underlying
distribution of relative orientations \citep{schmitt}. 
The notion that disks are intrinsically perpendicular to jets \citep{kleijn}
may also be an artifact of observational biases and small sample sizes, or 
because only a restricted class of objects 
exhibit jet/disk orthogonality. 

\begin{figure*}
\plottwo{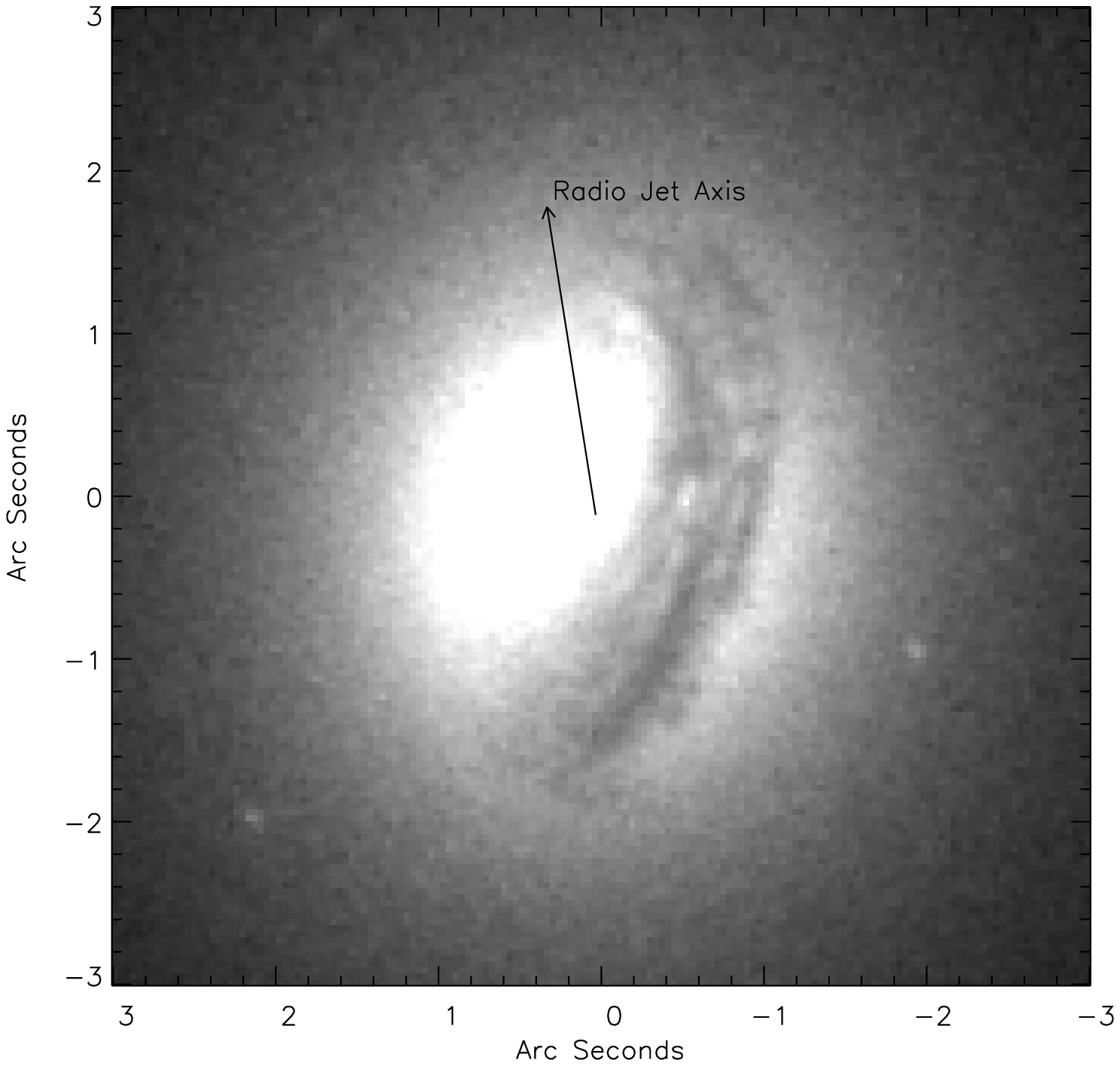}{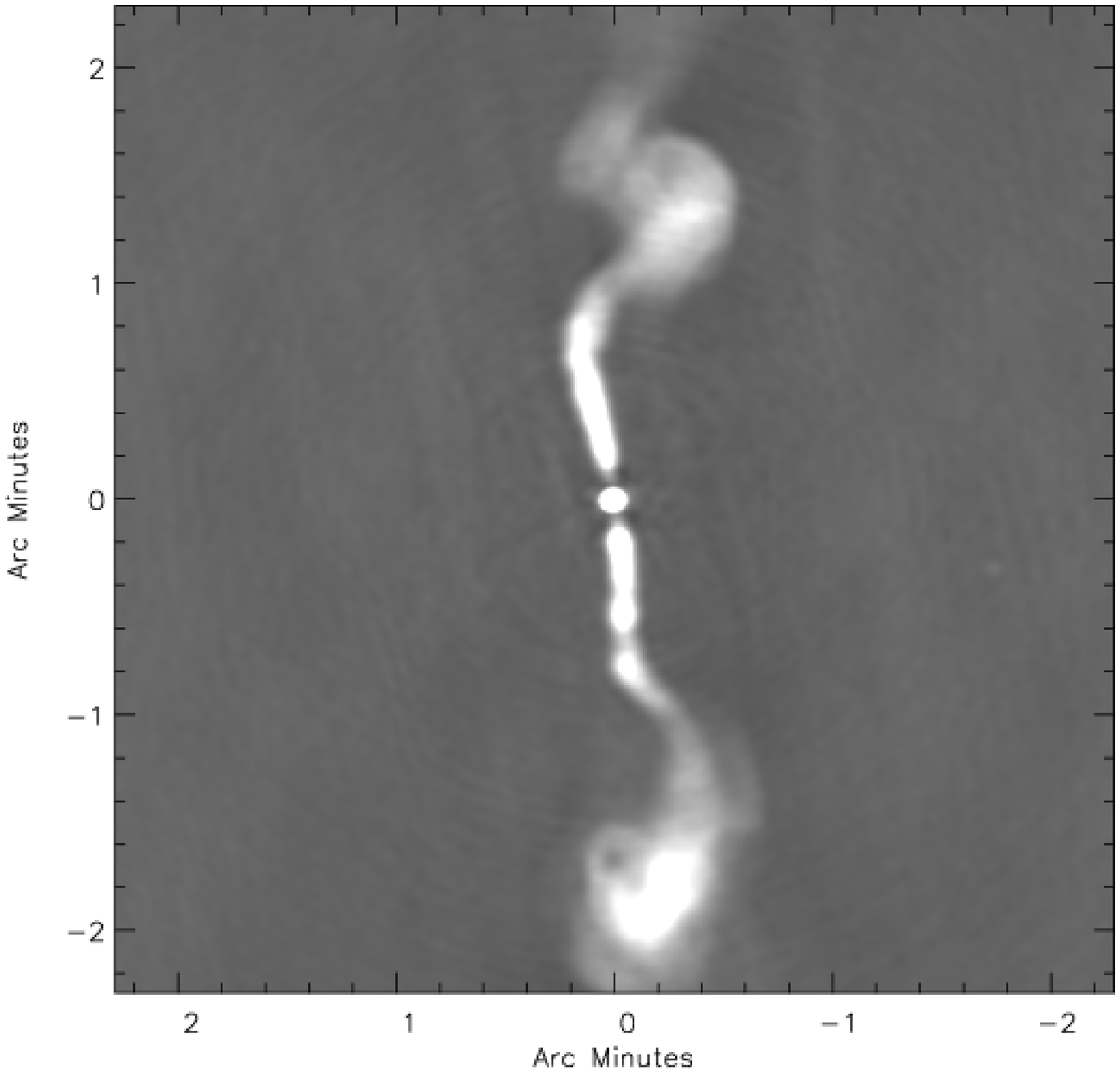}
\caption{
a) {\it HST/ WFPC2} 0.7 $\mu$m image of 3C 449.
Here the highly inclined nuclear disk is clearly visible.
b) 4.8 GHz {\it VLA} image of the radio structure of 3C 449.
At a distance of 73 Mpc, one arcminute
corresponds to $\sim 21$ kpc.}
\label{fig:jetdisk}
\end{figure*}

An additional complication in the interpretation of any $\Psi_{JD}$ distribution
arises from the sometimes disturbed morphology
of gaseous structures in the heart of 
many radio galaxies (e.g., \citealt{martel00, dekoff00,vandokkum95}).
Such studies suggest that many of these disks are not 
dynamically stable or relaxed in the galactic potential.
Detailed observations of gaseous structure in individual galaxies
have also found evidence for non-planar or warped geometries 
(e.g., \citealt{dekoff00,vandokkum95},
\citealt{ferrarese99} in NGC 6251 and \citealt{quillen99} in M84).
Furthermore, \citet{noelstorr05} presents an emission-line study of gas kinematics 
in the heart of radio galaxies, finding that many such structures 
are not consistent with a thin Keplerian disk model. 

In a non-spherical quiescent galaxy, dust and gas settle into 
a galactic symmetry plane on timescales of order 10-100 times 
the rotation period at a given radius \citep{ste88}.
Previous studies have shown that the dust in elliptical galaxies
is likely external in origin, and settles into the galactic equatorial plane
after several dynamical timescales following a merger
(e.g., \citealt{tran01,dekoff00,goudfrooij95,lauer05}).
Gas that does not reside in the principle symmetry plane of a galaxy 
must have been injected or perturbed on a timescale 
shorter than this.   
Since the rotation period is short near the nucleus
(approximately 2 million years at 100 pc in a luminous elliptical galaxy),
we would expect to find the majority of
nuclear gas disks residing in a symmetry plane of their host galaxies.
As such, we would not expect a link between the orientations of
$\sim 10^2$ pc sized gas disks and radio jets.
Instead, we generally expect to find settled, planar, and quiescent
nuclear disks in the symmetry plane of the galaxy.
Nevertheless, disturbed dust morphologies are common in the
heart of radio galaxies,
and the observed correlation between jet and disk axes 
requires a physical connection between these structures over either a small
range of orientations (e.g., the cone of avoidance model) or for a subclass of 
radio galaxies.   

Relativistic effects are known to perturb the inner accretion 
region, but the spin of a massive Kerr black hole can only 
strongly affect
the orientation of a disk within about ten thousand times 
the gravitational radius, $r_g = GM_{bh}/c^2 \sim 10^{14}$ cm,
where $M_{bh}$ is the mass of the black hole. 
Outside a few parsecs, Lense-Thirring precession
and the associated settling into the midplane, called the
Bardeen-Petterson effect, requires longer than a Hubble time to operate 
\citep{lense, bardeen75, caproni, kumar}.
The precession rate is 
\begin{eqnarray}
\Omega_{LT} &=& { 2 a G^2 M_{bh}^2  \over c^3 r^3} \\
\qquad &=& 7 \times 10^{-16} {\rm yr}^{-1}
 \left({a \over 0.5}\right)
 \left({M_{bh} \over 10^9M_\odot} \right)^2
 \left({r \over 100 {\rm pc} }\right)^{-3} \nonumber
\label{eqn:LT}
\end{eqnarray}
where $a$ is the dimensionless spin parameter.
At large radii, 
the disk is not expected to align with the equatorial plane of the black hole. 
The inner edge of the disk near $10 r_g$ 
is the proposed site for jet collimation and acceleration
\citep{ree82}, and most studies predict that the jet 
should be aligned with the spin axis 
of the black hole (e.g., \citealt{bardeen75,kumar}) but
not necessarily the angular momentum of the disk well outside of $r_g$.
A massive spinning black hole can be described as an angular momentum 
reservoir that varies in momentum, but only very slowly 
with the influx of fuel.
The rate of exchange of angular momentum between
the outer disk and black hole is expected to be particularly
slow ($\gtrsim 10^9$ years) in radio galaxies where the black holes 
are massive ($\sim 10^9 M_\odot$)
and the fueling rates are low ($\sim 10^{-4} M_\odot$~yr$^{-1}$, 
or well below the Eddington rate) \citep{rees78}.
Thus, the angular momentum of the outer regions of the disk, hundreds 
of parsecs away from the black hole, is not expected to play a role 
in aligning the jet axis.
The apparent statistical bias toward orthogonal jet/disk orientations 
therefore presents a mystery.  
We expect the angular momentum of the very {\it inner} regions of the disk 
to couple with the spin of the black hole and the jet axis, but currently there 
is no consistent mechanism capable of accounting for jet/disk alignment
on 10$^2$ pc scales.

Noteworthy exceptions to the above generalizations
have been considered by a number of studies. 
A merger of two black holes
could change the spin axis of the active galactic nucleus
(AGN) on a very short timescale \citep{merritt,liu04}. 
\citet{lubow} showed that when $\beta< h/r$, bending waves may
cause the inner disk to be misaligned with the black hole equatorial plane. 
(Here, $\beta$ is a dimensionless viscocity parameter, $h$ is the 
Gaussian scale height, and $r$ is the radius of the disk).
\citet{natarajan98,scheuer} showed that the timescale
for a black hole to align with a disk should depend on 
the ratio between the black hole and disk angular momentum times the
Lense-Thirring precession time, rather than an accretion timescale.
\citet{quillen99,quillen97} showed that the pressure
in the ambient interstellar medium (ISM) is sufficiently large to 
perturb a dusty disk, causing it to precess about the jet axis.   
Structure associated
with radio jets has been clearly revealed in recent observations
of the X-ray emitting gas in radio galaxies 
(e.g., \citealt{fabian03,hardcastle98}), implying that there could be 
a connection between nuclear activity and the energetics
of the ISM in radio galaxies and galaxy clusters.
The radiative instability proposed by \citet{pringle} could also
cause a disk to warp (see also \citealt{maloney96}).

\subsection{The nuclear disk of 3C~449}

It is useful to consider radio galaxies for which 
the jet and disk axes are clearly not orthogonal or are near
the cone of avoidance estimated by \citet{schmitt}.
Specimens possessing extreme (nearly parallel) 
jet/disk orientations serve as important test cases
for physical models capable of coupling
the energetics of radio and gaseous structures.

To that end, this paper presents a study of the radio galaxy 3C 449.
The elliptical galaxy UGC 12064 is the low redshift ($z=0.0171$)
host to this FR I type radio source,
which was imaged with the {\it Hubble Space Telescope} ({\it HST}) 
{\it WFPC2}
camera in the F702W filter centered at 0.7 $\mu$m  \citep{dekoff00}.
Figure \ref{fig:jetdisk} (a) reveals 
a large,  $\sim 600$ pc dusty gas disk
with faint spiral features, lying in a plane that is
is nearly parallel to the jet axis, 
as seen in Fig.~\ref{fig:jetdisk} (b).
Projected on the sky, the jet axis is at a position angle of 
$PA \approx 9^\circ$  \citep{feretti99},
running nearly north to south,  whereas the major axis of the disk
has $PA \approx 160^\circ$ \citep{dekoff00}. 
As such, the jet/disk angular difference
$\Psi_{JD} \approx 29^\circ$, establishing 3C 449 as 
a counterexample to the jet/disk orthogonality trend proposed by \citet{dekoff00},
being the only FR I radio galaxy
(out of 9) to possess dust features with
$\Psi_{JD} < 50^\circ$. In the sample discussed by \citet{kleijn},
6 out of 33 radio galaxies with measured jet/disk position angles
have $\Psi_{JD} \lesssim 50^\circ$.  
For 3C 449, the angle made by the rotational axis 
of the disk and the jet axis is greater than
$60^\circ$, placing 3C 449 near the cone of avoidance estimated 
by \citet{schmitt}.
3C 449 is at a distance of 73 Mpc, 
such that $1\arcsec$ corresponds to 350 pc for a Hubble constant 
of 70 km~s$^{-1}$~Mpc$^{-1}$.
The maximum angular resolution of the {\it HST} data is 
$\sim 0\arcsec.1$, corresponding to $\sim 35$ pc.
The mass of the black hole residing at the nucleus of 3C 449 is estimated
at $2.5 \times 10^8 M_\odot$ \citep{bettoni03}.
Based on measurements of the central core at optical, radio
and X-ray wavelengths, the active nucleus of 3C 449 
has a bolometric luminosity $\sim 10^{42}$~erg~s$^{-1}$ or 
$\sim 6 \times 10^{-4}$ of the Eddington luminosity \citep{donato}.

\begin{figure*}
\plotone{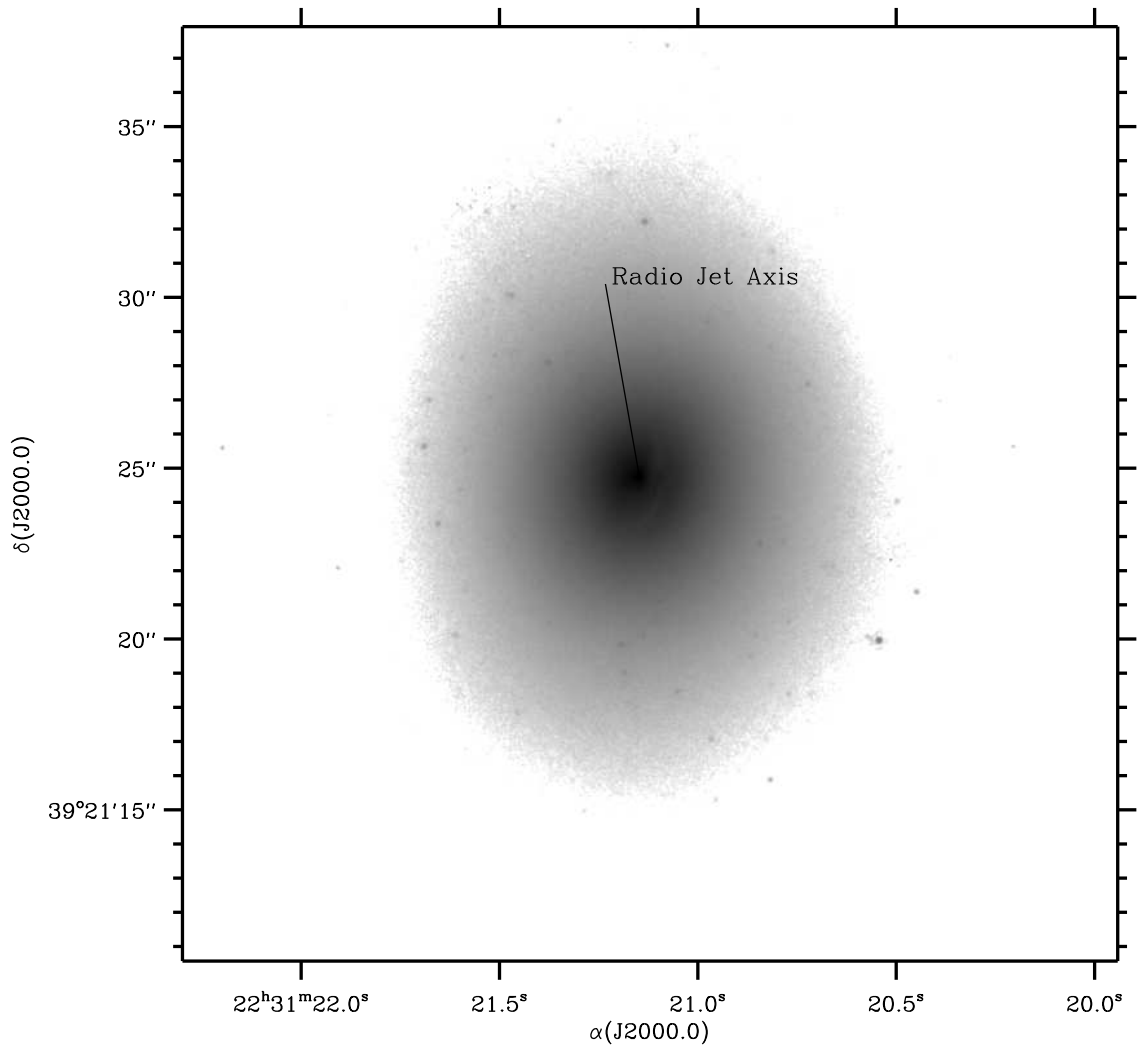}
\caption{
{\it HST/NICMOS} 1.6 $\mu$m image of 3C 449.
The near side of a dusty torus is seen in absorption
on the western side of the nucleus. The jet axis is nearly
in the plane of the sky at $PA=9^\circ$.
}
\label{fig:sky}
\end{figure*}

The recent {\it HST/NICMOS} 3CR snapshot survey \citep{snap}
allows for a more detailed study of the dust features in UGC 12064.
Because extinction from dust is reduced at 1.6 $\mu$m compared
to visible wavelengths, the {\it NICMOS} image provides
a less obscured view of the underlying 
stellar surface brightness profile.
By combining this near-infrared data with the 0.7 $\mu$m image, we
created a colormap which highlights
the absorptive features and isocolor contours (isochromes) of the disk. 
We describe these features in \S2, and discuss an observed 
twist in the isochromes that is inconsistent
with the absorptive properties of a planar disk.
In \S3 we describe our method of creating model images by integrating 
galactic starlight through an absorptive tilted-ring disk model.   
We explore non-planar structures for the model disk, finding that
a warped geometry provides a possible 
explanation for the peculiarities observed in the colormap.
Our model is used to constrain the morphology of the dusty disk, and 
in \S4 we discuss physical mechanisms which
could account for the warp predicted by our model.
A summary and discussion follows.

\begin{figure*}
\plotone{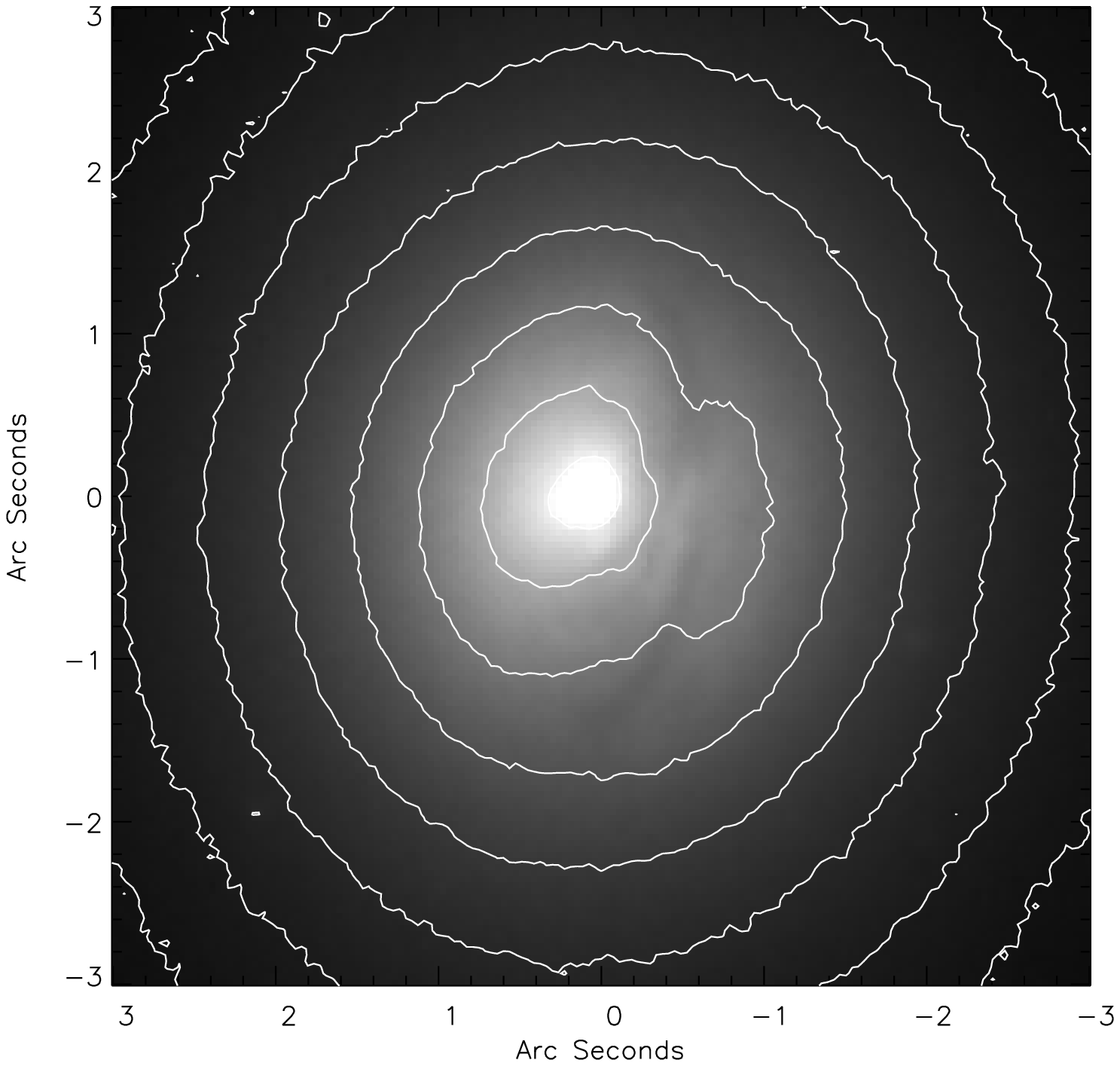}
\caption{
{\it HST/NICMOS} 1.6 $\mu$m image of 3C 449 with contours
marking isophotes.
The contours highlight every 0.4 magnitude difference in brightness, moving
outward from the core.
The innermost contour marks a brightness of 13.7
$H$-band magnitudes per square arcsecond.
Note that the major axis of each
contour is generally aligned with all others, suggesting that
the galaxy is unlikely to be triaxial.
}
\label{fig:nic}
\end{figure*}

\section{Observations}

\begin{figure*}
\plotone{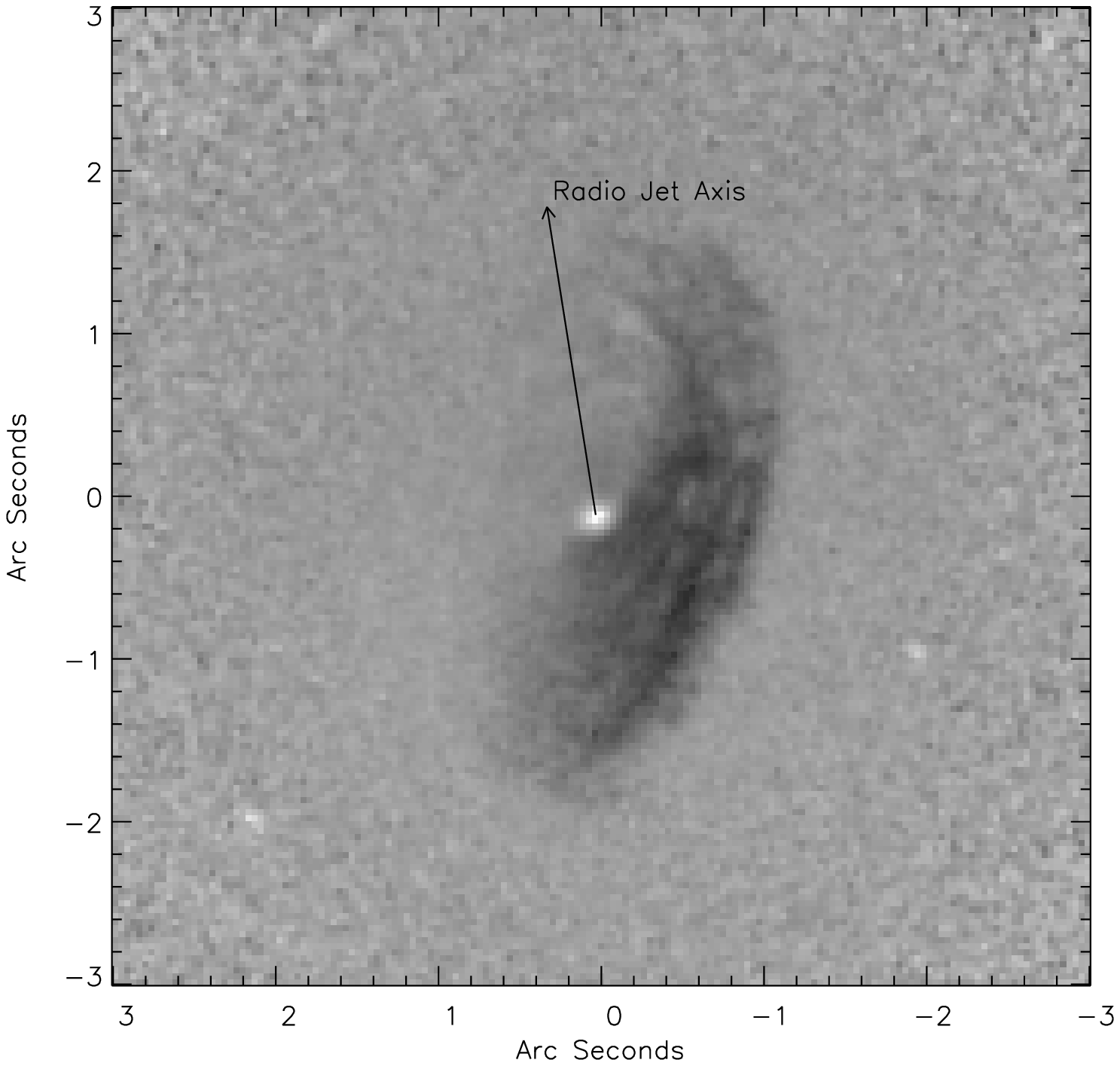}
\caption{
Colormap of 3C 449 generated via division of the 1.6 $\mu$m {\it HST
NICMOS} image by the {\it WFPC2} image at 0.7 $\mu$m.
Note the reddening of the disk
south of the nucleus. There is approximately a 0.5 magnitude difference
between the faintest and brightest portions of the colormap.
The radio jet axis is at a position angle of $9^\circ$, measured north through east.
Note the misalignment of the isochromes, particularly the
``integral-sign'' like twist
in those bordering the southwestern side of the nucleus.
A planar disk should exhibit isochromes that are aligned
with the disk major axis at all radii, which is clearly not the case for this object.
}
\label{fig:color}
\end{figure*}

3C 449 has been observed as part of a near-infrared snapshot survey carried
out with the {\it NICMOS} 2 camera on board {\it HST}, using the F160W
broad band filter at 1.6 $\mu$m \citep{snap}.
After aligning and scaling the images, 
we created a colormap via division of this data with
the 0.7 $\mu$m image taken with {\it WFPC2} and
previously studied by \citet{martel00,dekoff00}.
PSF differences between the two images were ignored, as any variation 
would only strongly affect regions of high surface brightness, (e.g.~near the nucleus). 
The images and colormap are shown in Figures \ref{fig:jetdisk}(a),
\ref{fig:sky}, and \ref{fig:color}, respectively.
A highly inclined, sharply edged disk can be seen in both the 
0.7 and 1.6 $\mu$m images of Fig.~\ref{fig:jetdisk}(a) and \ref{fig:sky} and the corresponding colormap
(Fig.~\ref{fig:color}); a structure previously described by 
\citet{martel00,dekoff00}.
The 0.7 $\mu$m {\it WFPC2} image reveals more filamentary dust structure
than the near-infrared 1.6 $\mu$m {\it NICMOS} image.
We estimate 
that the $\sim 600$ pc disk has an axis ratio of $\sim 0.56$ near its 
outer edge, with the $3\arcsec.45$ long major 
axis lying at position angle $PA \sim -20^\circ$.

As expected, absorption from the disk is clearly less prominent
at 1.6 $\mu$m than at 0.7 $\mu$m, so the 1.6 $\mu$m 
{\it NICMOS} image was used to fit a 
S\'{e}rsic surface brightness profile to the galaxy.
We use this profile to represent the underlying stellar
distribution in our models (see \S3).
The isophotes at 1.6 $\mu$m in the central region have ellipticity
ranging from 0.1 at a radius of $2\arcsec$ to 0.2 at $13\arcsec$.  
As can be seen from the contours in Fig.~\ref{fig:nic},
there is no significant twist in the isophote orientation, 
as each isophote is aligned with major axis at $PA \approx -10^\circ$.

The major axes of nuclear disks in normal elliptical \citep{tran01} and 3C radio
elliptical galaxies \citep{martel00} tend to be aligned
within $15^\circ$ of the galaxy isophotal major axis, and 3C 449 is no exception. 
The galactic isophotal axis, lying along $\sim -10^\circ$ at radii just outside
the disk (of radius $\sim 2\arcsec$), is within $10^\circ$ of 
the major axis of the outer edge of the disk, at $\sim - 20^\circ$.
This suggests that the outer regions of this disk could be 
dynamically relaxed in the galactic potential.

A closer look at the colormap in
Fig.~\ref{fig:color} reveals the surprising isochromal properties of the disk. 
Note how it reddens dramatically 
from north of the nucleus along the major axis (at $PA\sim 160^\circ$)  
to the southwestern side, below the nucleus.
Even more surprising is the ``integral-sign'' like twist
in the isochromes bordering the southwestern edge of the nucleus.
This is unexpected, as a
planar absorbing disk should possess elliptical isocolor contours 
that are aligned with the disk major axis
and exhibit reflective symmetry about the disk minor axis. 
This is not the case in the colormap of 3C 449.

Note the linear streaks along the disk in the colormap, 
which have been described as spiral arms or dusty filaments by previous
studies \citep{dekoff00,martel00}. 
Near the nucleus, linear dust filaments are tilted 
at an angle more nearly east-west than in the outer disk. 
At a radius of $0\arcsec.2$ the linear dust features
are at a shallow position angle of $\sim 110^\circ$, whereas near the edge
of the disk, the features are nearly aligned with the disk
edge at $PA \sim 160^\circ$.
For a planar disk, one would expect tightly wound spiral features 
to cause dark absorption lanes that are approximately 
parallel to the outer disk edge at all radii. 
However, if the spiral spiral features are more open (less
tightly wound) near the nucleus they may cause a shift in
the observed position angles of the features.

The isophotes in the 1.6 $\mu$m image of Fig.~\ref{fig:nic} are nearly round
in front of the disk (to the northeast), so there is no evidence that 
the stellar distribution is lop-sided or barred. 
The disk outer edge has nearly the same major axis as the galaxy isophotes. 
Outside regions where there is strong extinction from the disk, we 
see no significant twist in the position 
angle of the galaxy isophotes as a function of radius.
Thus, we have no reason
to suspect that the galaxy is strongly triaxial in its central kpc-scale regions
A stellar bar with a major axis of $3\arcsec$-$4\arcsec$ would be capable of causing 
a twist in colormap isochromes, 
but a bar that is sufficiently triaxial to account for
such a twist would likely need to rotate to be stable.
A rotating bar generally truncates sharply near its corotation radius.
The surface brightness profile 
on the northeast side is smooth (and well fit
by a S\'ersic profile), so it is unlikely
that the stellar distribution is strongly triaxial (rotating or otherwise) 
in the central few arcseconds from the nucleus. 
As such, the twist observed in the colormap is most likely caused by the disk 
itself, rather than a projection effect from a strongly
triaxial nuclear cusp. 

This motivates the creation of an absorptive numerical model 
for the disk and underlying starlight distribution. 
In \S3 we explore warped geometries for the disk in an attempt 
to reproduce the isochrome twist seen in Fig.~\ref{fig:color}, and constrain
the geometry of the disk accordingly.  

\section{A Numerical Warped Disk Model for 3C 449}

We now describe our procedure for creating an absorptive
tilted-ring model of the gas disk of 3C 449. 
The starlight from behind the disk is integrated, multiplied by
an absorption factor due to the disk 
and added to the integrated starlight
from in front of the disk.
To create such a model, we require a light density function and
a geometric model for the disk which specifies
its opacity as a function of position on the sky and its
position along the line of sight in the galaxy.

We describe the stellar luminosity density 
$\rho(s)$ where $s = \sqrt{(1-e_g^2) r^2 + z^2}$. 
The ellipticity of the galaxy,
$e_g$, was not allowed to vary with radius.
Our function for $\rho(s)$ is consistent with the S\'{e}rsic law, 
fit to the surface brightness profile of the {\it NICMOS} image
using the routine {\it GALFIT} \citep{peng}.
The galactic surface brightness profile is
\begin{equation}
I(s)=\exp \left[{ -b_n \left({s\over s_0}\right)^{1\over n}} \right]
\label{eqn:brightness}
\end{equation}
where the scale length $s_0$ and power $n$ are listed in Table 
\ref{tab:tab1}; (for $n=4$ a de Vaucouleur law is recovered). 
Here $b_n = 2n-{1\over 3}+{0.009876\over n}$, 
as approximated by \citet{prugniel}.

The light density from stars (in 3 dimensions) that
is consistent with the S\'{e}rsic surface brightness profile
is given by \citet{prugniel},
\begin{equation}
\rho(s) = s^{-\alpha_r} \exp 
               \left[ {-b_n \left({s\over s_0}\right)^{1\over n}} \right]
\label{eqn:density}
\end{equation}
where $\alpha_r$ is a free parameter well approximated with the expansion
$\alpha_r = 1.0-{1.188 \over 2 n}+{ 0.22 \over 4 n^2}$.
We checked that this density function provides a good
match to the surface brightness profile by comparing 
the model light, generated from Eq.~\ref{eqn:density} and integrated 
along the line of sight, 
with the surface brightness
profile at 1.6 $\mu$m and the S\'{e}rsic function that was fit to it.

The dusty disk is assumed to be comprised of a series of rings 
of radius $r$, each with an orientation described by two parameters: 
a tilt with respect to the line of sight, $\omega(r)$, and $\alpha(r)$,
the position angle from north on the sky. 
We have defined these angles with respect to the line of sight,
not with respect to the galactic midplane or the jet axis.
We assume that $\omega(r)$ and $\alpha(r)$ are powers of $r$, such that
$\omega(r) \propto r^{a_\omega}$ and $\alpha(r) \propto r^{a_\alpha}$.
The opacity of the disk (if viewed face-on) is described by
$\tau(r) \propto r^{a_\tau}$. 
We specify the angles and opacity at two radii, $r_{min}$ and $r_{max}$. 
We assume that interior and exterior to these radii the disk opacity is zero.

We create two arrays, with coordinates $x,y$ corresponding
to positions on the sky.  The first array contains the opacity
of the disk at this position; the second contains
the location of the disk along the line of sight.
Our numerical procedure begins with a random selection of points
$r,\theta$ in the disk.  
Using the ring 
orientation functions $\alpha(r)$ and $\omega(r)$, the position
of this point is projected
to determine $x,y$ on the sky and $z$ along the line of sight.
The disk opacity is computed from the radius, and stored along with 
the vertical position of the disk in the two arrays described above.

At each position on the sky, light from the galaxy is integrated
using the stellar luminosity density $\rho(s)$ from $- \infty$
to the disk surface (with position stored in the second array). 
The sum of light from behind the disk is multiplied by
$e^{-\tau}$ at that position (from the other array)
to take into account absorption by the disk.
This flux is then added to the integrated star light in front of the disk
(from $+ \infty$ to the disk surface).
The opacity at 0.7 $\mu$m of the dusty disk 
is assumed to be 4.16 times that at 1.6 $\mu$m,
consistent with a galactic extinction law \citep{mathis90}.
The output of the light density integration produces model images
at 0.7 $\mu$m and 1.6 $\mu$m. As was the case for the {\it HST} data, we create
a model colormap via division of these two output images.

\subsection{Comparison of model and galaxy images}

\begin{figure*}
\plottwo{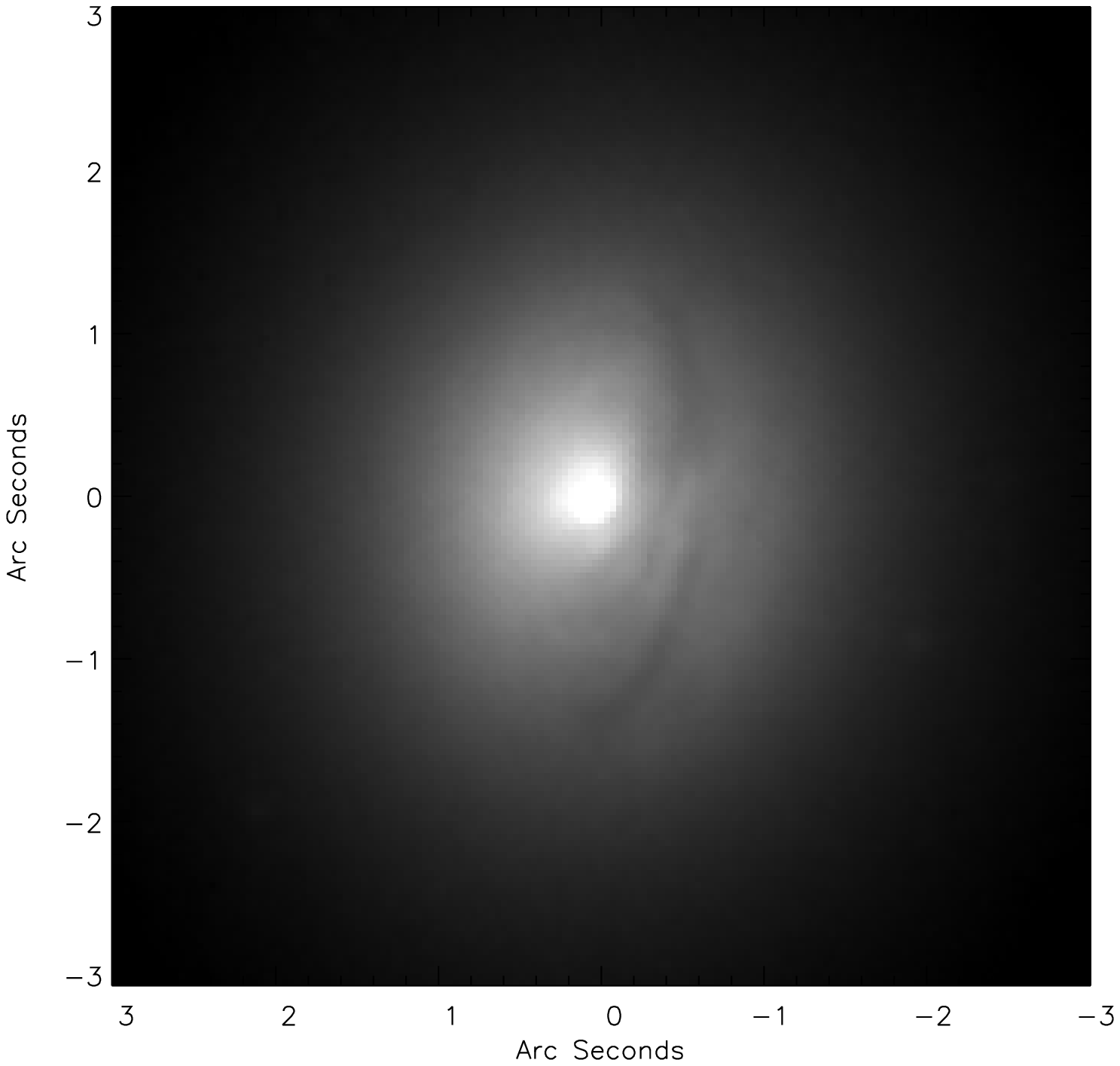}{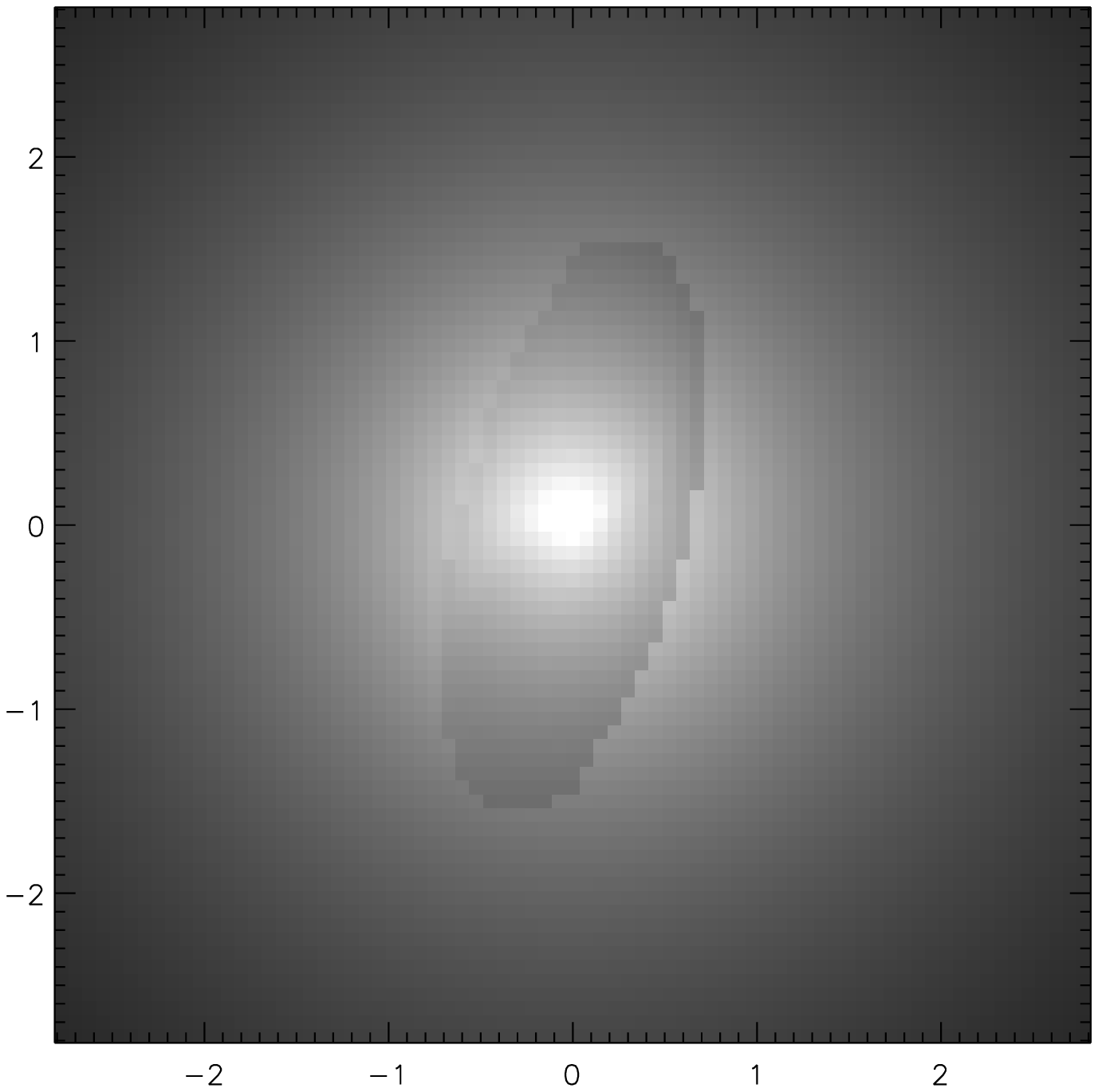}
\caption{
Comparison of a)  1.6 $\mu$m {\it  NICMOS} image with
b) the 1.6 $\mu$m integrated light warped disk model.
Note the similar absorptive properties
of both the disk and its numerical representation.
The disk in the model image appears larger than that in the {\it NICMOS}
data as extinction from the galaxy masks the outer regions of the disk
at 1.6 $\mu$m. In our model, we do not take into account extinction from
the surrounding galaxy.
The integrated light density function used in creating the model is based
upon isophotal analysis of the {\it NICMOS} image.
The $R$-band opacity drop-off $\sim 0.05$  follows a simple
power law in an attempt to obtain a qualitative match to the data (see \S3).
Model parameters are listed in Table \ref{tab:tab1}.
}
\label{fig:nic_compare}
\end{figure*}

\begin{figure*}
\plottwo{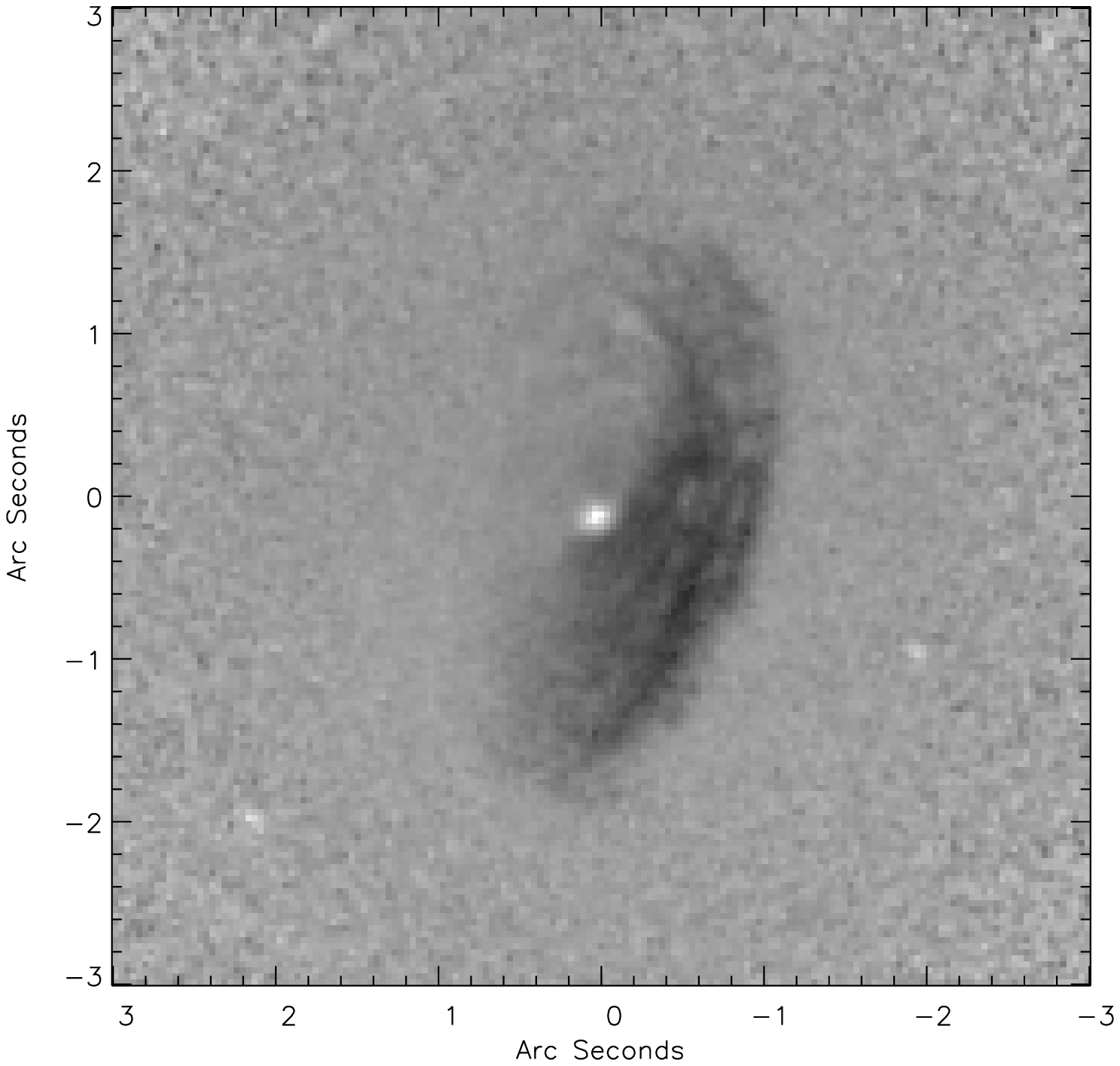}{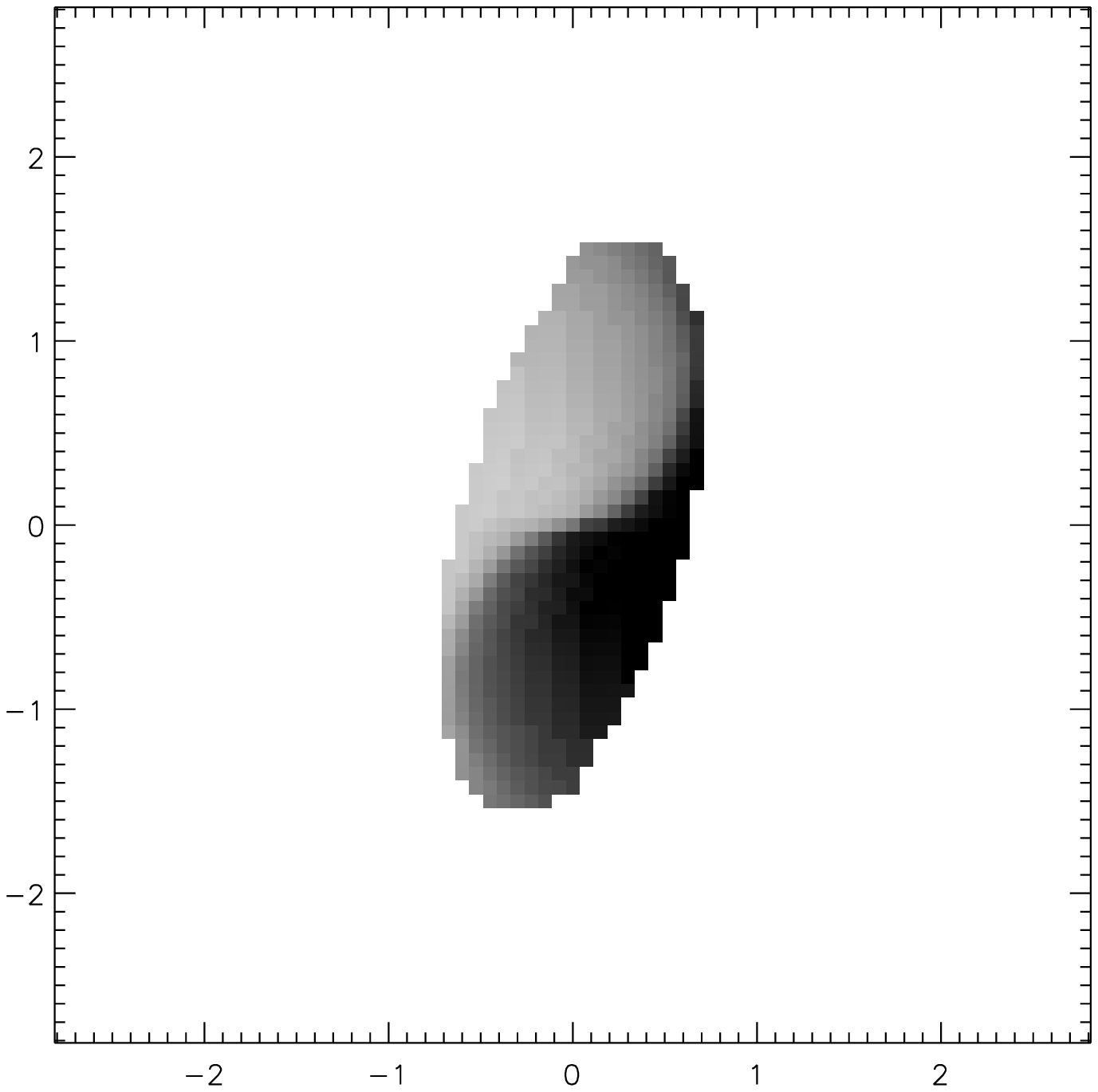}
\caption{Comparison of a) colormap of 3C 449, generated
via division of the 1.6 $\mu$m and 0.7 $\mu$m {\it HST} images, with
b) the model colormap.
Note the successful reproduction of the isochrome twist to the north
of the nucleus.
Our model assumes that the disk is warped but featureless,
so the spiral features are not seen in the model.
Model parameters are listed in Table \ref{tab:tab1}.
}
\label{fig:color_compare}
\end{figure*}

Our goal in creating models is to test the hypothesis 
that the disk of 3C 449 is warped, and to better 
constrain the geometry of the disk. 
We alter model parameters to obtain a qualitative
match to the morphology of both the colormap and the 1.6 $\mu$m image.
Our best matching model, shown in Fig.~\ref{fig:nic_compare},
compares the image at 1.6 $\mu$m to the model 
image, and Fig.~\ref{fig:color_compare} shows the colormap compared
to the model as viewed in 1.6 $\mu$m/0.7 $\mu$m.
The parameters for this model are listed in Table \ref{tab:tab1}.

\begin{deluxetable}{lrr}
\tablewidth{0pt}
\tablecolumns{3}
\tablecaption{Parameters for Warp Model \label{table:tab1}}
\tablehead{
\colhead{Parameter} &
\colhead{Value} &
\colhead{Comments}
}
\startdata
\cutinhead{Isophotal Parameters}
$n$             &  1.2797    & From {\it GALFIT}\\
$s_0$           &  $5\arcsec.6625$& From {\it GALFIT}\\
$e_g$           &   0.175    & From {\it GALFIT}\\
\cutinhead{Opacity Parameters}
$r_{min}$       &  $0\arcsec.1$   & Minimum {\it HST} resolution\\
$r_{max}$       &  $1\arcsec.6$   & Estimated from Fig.~\ref{fig:color}\\
$\tau_{min}$    &  0.04      & Estimated opacity at $r_{min}$\\
$\tau_{max}$    &  0.09      & Estimated opacity at $r_{max}$\\
$a_\tau$        &  1.0       & For $\tau(r) \propto r^{a_\tau}$\\
\cutinhead{Warp Parameters }
$\omega_{min}$  & $-10^\circ$& Disk inclination at $r_{min}$\\
$\omega_{max}$  & $-23^\circ$& Disk inclination at $r_{max}$\\
$a_\omega$      &   1.0      & For $\omega(r) \propto r^{a_\omega}$\\
$\alpha_{min}$  & $-73^\circ$& $PA$ of disk at $r_{min}$\\
$\alpha_{max}$  & $-14^\circ$& $PA$ of disk at $r_{max}$\\
$a_\alpha$      &   1.0      & For $\alpha(r) \propto r^{a_\alpha}$\\
\enddata
\tablecomments{
Isophotal parameters are from {\it GALFIT} \citep{peng}.
Angles for $\alpha$ are position angles ({\it PA}), measured from north
through east on the sky (counterclockwise).
$\omega$ is a tilt with respect to the line of sight ($\omega=0^\circ$ for
edge on disk). A negative angle for $\omega$ means that the southern edge
of the disk is the near side.
The jet axis is assumed to be in the plane of the
sky and at $PA=9^\circ$.
}
\label{tab:tab1}
\end{deluxetable}

Figures \ref{fig:nic_compare} and \ref{fig:color_compare} illustrate
that a warped disk model is successful in reproducing the 
appearance of the {\it HST} images. Specifically, we find that
absorption in a nonplanar disk can give rise to an ``integral-sign'' like 
twist in isochromes, as is observed in the colormap of 3C 449.  
The model also successfully exhibits
absorption along the western edge of the disk in the 1.6 $\mu$m image.
In $H$-band, light from behind the disk is absorbed
by the southwestern (nearest) side, but is visible
past the edge. The same is true for the representative
model, which differs from the {\it NICMOS} image in that
the far side of the disk is also visible and appears larger.
This is due to extinction effects leading to obscuration
of the disk at 1.6 $\mu$m. The colormap reveals the
underlying morphology of the absorptive dust features, so it was used
as the basis for shaping our model disk.
Because our model lacks azimuthal (non-radial) density
variations or spiral surface features,
it cannot reproduce the linear streaks seen in
absorption in the 1.6 $\mu$m {\it NICMOS} image.

Fig.~\ref{fig:color_compare} is more convincing, as it shows that
our model can reproduce the ``integral sign''
twist in the isocolor contours near the nucleus of the colormap, as well as 
match the shape of the absorptive features quite well, 
particularly on the southwestern edge of the disk.   

Certainty and degeneracy in model parameters is of key 
importance in judging the qualitative success of the model.
We first list the parameters that were held fixed and then
describe those that we varied to find a model that best
matched the images.
The isophotal parameters $n$ and $s_0$ were fixed since
they are consistent with the S\'{e}rsic law 
fit to the surface brightness profile of the galaxy. 
We fixed the galaxy ellipticity $e_g$ at 
0.175, as this produces model isophotes that are sufficiently
oblate to be consistent with the 
isophotal ellipticity at 1.6 $\mu$m. 
We set the maximum radius of the disk to match that of the outer
disk edge, $r_{max} = 1\arcsec.6$.   Likewise the tilt, 
$\omega_{max} =-23^\circ$ (w.r.t. the line of sight), and position angle, $\alpha_{max}=-14^\circ$,
of the outermost ring (at radius $r_{max}$) were fixed and
set to be consistent with the location of the disk's outer edge. 
The inner model ring was set with $r_{min}=0\arcsec.1$, corresponding to 
the maximum angular resolution of {\it HST}. We cannot probe the 
geometry of the disk at radii smaller than this, so 
this radius does not describe an inner edge for the disk.  
Rather, it is used to provide boundary conditions for the positional 
functions $\alpha(r)$ and $\omega(r)$, and for the opacity function $\tau(r)$.

The model is strongly
sensitive to changes in the optical depth of the disk, 
which is set by the parameters
$\tau_{min}$ and $\tau_{max}$ at the innermost and outermost rings, and 
with an exponent, $a_\tau$.
Such sensitivity allows us to constrain the opacity parameters 
by incrementally altering $\tau_{min}$ and $\tau_{max}$ 
until the isophotes of the model 
colormap match the images.
Too high an opacity leads to overly strong absorptive features 
not observed in the {\it NICMOS} image, 
and opacity endpoints that are too low have the opposite effect. 
The best match for our model was in the range
of $R$-band opacity $\tau_{min} = 0.04 \pm 0.02$ to $\tau_{max} = 0.09 \pm 0.02$.
This physically reasonable $R$-band opacity range
generates isochromes that match those seen
in the {\it HST} colormap.  
Still, there is inherent uncertainty in our values for these parameters, 
as their effects on the absorptive features differ depending on the shape of the 
disk. A disk of constant density has a higher opacity when viewed
at high inclination than when viewed face-on.
Thus, there is a larger range of values over which the inner ring
opacity will produce an ``integral-sign'' twist. This, along 
with the orientation of the inner ring, is the greatest source of uncertainty 
in our model.

The model is not as sensitive to the exponent $a_\tau$ 
in the power law $\tau(r) \propto r^{a_\tau}$. 
While opacity and ring orientation effect 
the structure of the model isochromes, $a_\tau$ effects how steeply an absorptive 
feature tapers off as a function of radius. 
This dependence is minimal, in that a model with $a_\tau = 0.5$ and 
a model with $a_\tau = 1.5$ will produce similar results, both of which 
closely resemble the data if the positional parameters and opacity boundary conditions 
are set appropriately.
As such, $a_\tau$ is only loosely constrained, and we set it to $a_\tau = 1.0$ as it 
generates absorptive features in the model colormap that are consistent with those 
seen in the data.

While the strength of the absorptive features is mostly dependent upon $\tau(r)$,
isochrome {\it shape} most strongly depends on the structure of the disk.
Qualitatively, one can achieve similar outer disk appearances by 
raising $\alpha$ and decreasing $\omega$ (or vice-versa). 
Being concerned only with the appearance of the disk, 
the small range over which $\alpha$ and $\omega$ are degenerate 
is of little relevance.  
The orientation of the innermost ring 
is of particular interest, since it is located only 
$\sim 30$ pc from the nucleus. 
The best matching model exhibits a dramatic warp, 
with the inner ring at a radius of $0\arcsec.1$ at a position angle
of $\alpha_{min}=PA= -73^\circ$. 
We estimate the best fit for the tilt of the inner ring is at 
$\omega_{min}=-10^\circ$ w.r.t. the line of sight.   

Remarkably, the best matching model is such that 
the inner regions of the disk are nearly perpendicular to the jet axis. 
Assuming the jet axis is orthogonal to the line of sight (e.g., in the plane 
of the sky) and at $PA=9^\circ$, the axis of symmetry of the innermost 
model ring makes a 3D cone angle with the jet axis of approximately $10^\circ$, meaning 
the inner regions of the disk are within $10^\circ$ of being orthogonal to the jet axis. 
The uncertainty brought about by the 
codependence of $\alpha$ and $\omega$ on disk appearance
is small enough such that we can be fairly certain of this value, as all ``good-fit'' 
models have inner rings nearly perpendicular to the jet axis. 
Consequently, our model shows that 3C 449 may actually {\it conform} 
to the trend proposed by previous studies
suggesting that dusty disk major axes are preferentially 
oriented nearly perpendicular to jet axes on the sky
\citep{dekoff00,vandokkum95,sparks}.

\subsection{Comparison of warp model to observations
at other wavelengths}

The success of the non-planar model in reproducing the observed
isochrome twist in 3C 449 leads us to suspect
that the dust disk is warped between a radius
of $\sim 50$ pc (the innermost radius we can easily resolve) 
and 600 pc (the disk outer edge). 
We now discuss the geometry and orientation of the dusty
disk in context with observations of the nuclear region at 
other wavelengths.

\citet{martel00} presented narrow band images of 3C 449 in H$\alpha$+N[II], 
showing stronger emission to the north of the nucleus.
Likewise, the near-UV image presented by \citet{allen02} 
showed similar morphology 
with brighter near-UV emission to the north of the nucleus.
The UV image exhibits arm-like features that are bounded
by dusty absorptive streaks also present in the visible band image
(as discussed by \citealt{allen02}).

Because extinction is deep along the disk edge to the west of the
nucleus, we infer that the western side of the disk is nearer to us than
the eastern side.  Studies of the radio emission have suggested
that the jets are nearly in the plane of the sky, 
at approximately $70^\circ$ from the line of sight, 
with the southern side nearer the observer \citep{gower82}.
As such, we expect that the northern jet is in front of the disk  
and the southern jet is behind.
The UV emission might be excited by star formation in the disk 
(e.g., \citealt{odea01}, in the case of 3C 236), 
or it might be related to shocks or illumination
along the jet axis (e.g., \citealt{solrzano}).
Since bright UV or H$\alpha$+[NII] emission is not seen on
the southern side or along the western disk edge, 
the emission probably does not arise from
material embedded in the disk.
If the emission is from a large diffuse
region (possibly wide cones) above and below
the disk and oriented approximately along the jet axis, 
the emission from southern side should suffer more 
extinction from the disk.   
Previous studies (e.g., \citealt{quillencone})
have found that structure in an ionization cone can be related
to the underlying dust distribution in the galaxy. 
However, the UV emission from 3C 449 is brighter outside
the dusty features than on top of them, though UV emission
is also brighter in proximity to dust features, so the presence
of dense material may somehow contribute to the excitation.
The morphology of the UV emission would be difficult
to account for with a simple model, but excitation along
a wide cone associated with the jet axis and coupled with
extinction from the dusty disk might be consistent
with the observed morphology, so long as the UV emission were stronger
in proximity to dust features.  The UV emission also
exhibits a deficit (or perhaps a shadow) just north of the nucleus. 
Our model does not include material that could
absorb UV emission in this location, but a warped
disk that continues to twist past our observational limit at $r_{min}$ could
possibly account for such a shadow.

We can consider the
possibility that blue continuum emission associated with the UV structure
could have reduced the redness of color to the north of the nucleus, 
accounting for the observed isochromal twist in the colormap.
If this were so, we would predict that the size and location
of features seen in the UV image would correspond to blue regions
in our color map.
The colormap is quite smooth, wherease the UV emission is clumpy.
Furthermore, bright UV emission is offset from the spiral features seen
in absorption.
This would make it impossible for a blue continuum associated
with the UV emission
to cancel out the extinction associated with dust lanes.

\subsection{Comparison to Spectra}

Our warp model predicts that the inner disk (at $r \sim 100$ pc or $0\arcsec.3$) 
is nearly edge-on
and oriented almost directly east-west, whereas the outer disk 
at $r \sim 600$pc is oriented with major axis oriented
approximately north-south.   
We can compare this
prediction to spectroscopic observations that measure the mean
velocity of the ionized gas.  
\citet{noelstorr} obtained spectra with {\it STIS} on board {\it HST}
along 3 slits oriented at a position angle of $170.6^\circ$ 
(approximately north-south along the major axis of the outer disk). 
Mean velocities measured along these slits are presented
in Figure 22 and Table 26 of \citet{noelstorr}.  
The mean velocity $\sim 0\arcsec.5$ north of the nucleus is 
$\sim 400$ km/s above that $\sim 0\arcsec.5$ south of the nucleus.
The observed velocity profile along the outer disk major axis  
is consistent with a rotating disk that is approaching the observer
on the northern side and receding on the southern side.
The warp model predicts that the eastern side of the
inner disk should have radial velocity with the same sign
as the southern side of the outer disk.
Because the southern side of the outer disk is receding from
us, our model implies that the eastern side of the inner disk
should be receding, while the western side 
should be approaching the observer.

Mean velocities were also measured in 
the two slits $\pm 0.\arcsec1$ from the nucleus and parallel 
to the central (nuclear) slit.  
Velocities along the east-west direction
can be used to test the sign of the rotation predicted by the model.
The velocities on either side (east and west) of
the nucleus are lower and higher (respectively) than the systemic velocity,
suggesting that the velocity field is twisted.
East of the nucleus, \citet{noelstorr} found the velocity 
to be sub-systemic. To the west, a velocity
consistent with that observed along the major axis south of the
nucleus was measured.   
This would be consistent with a disk that is rotating
such that southern and eastern sides of the
outer disk are receding from
the observer. We find that the sign of the velocities
along the east-west direction is consistent with the prediction
of our warped disk model.

\section{Mechanisms for Causing the Warp}

We have shown that a warped disk could account
for the twist in the isochromes seen in the dusty disk of 3C 449, 
and that our model is consistent with observations at 
other wavelengths.
With an outer edge at $r \sim 600$ pc, the disk 
is approximately in the midplane of the galaxy
and tilted about $30^\circ$ away from the jet axis.
If our model is correct, at smaller radii ($\sim 100$ pc),
the disk is nearly orthogonal to the jet.  
\citet{kleijn} outlined two scenarios that could account
for such a jet/disk orientation.
\begin{enumerate}
\item The disk is initially planar
and sets the jet axis.  At later times the disk is affected
by the torque from the galaxy and becomes aligned with the galaxy
isophotes. 
\item Another force, possibly aligned with the jets, pushes
on the disk, twisting it at small radii.
\end{enumerate}

Might the morphology of the disk in 3C 449 discriminate
between these two scenarios?  We can consider a gas disk
to be comprised of rings moving in nearly circular motion.  
When the galaxy is non-spherical, it exerts a torque on each ring
causing precession about a galactic axis of symmetry.
The precession rate is 
$\dot \alpha \sim \Omega \epsilon_\Phi$
where $\Omega$ is the angular rotation rate of a particle
in a circular orbit and $\epsilon_\Phi$ is the ellipticity
of the galactic gravitational potential. (Here $\alpha$
refers to a precession angle which is measured
with respect to a principal plane of the galaxy).   
The precession rate is primarily set by $\Omega$,
and so depends on the rotation period at a given radius,
$P = 2\pi/\Omega$. Since the rotation period is shorter
at smaller radii,
the inner disk should be more twisted than
the outer disk, unless the galaxy ellipticity drops rapidly near 
its inner regions. 
Such a drop in ellipticity is not seen in the isophotes
of Fig.~\ref{fig:nic},
though the presence of the disk itself does not
make this absolutely certain.  As the disk becomes increasingly twisted,
the inner region can settle into the galactic midplane
(e.g., \citealt{ste88}) with the outer disk warped 
and misaligned with the galaxy.  Instead, we find that
the {\it outer} part of the galaxy is settled into the 
midplane and the {\it inner} part deviates from
or is inclined with respect to this plane.  
This is not consistent with the first scenario
discussed above.  The morphology of the disk
instead suggests that it was once relaxed in the symmetry plane
and subsequently perturbed by a force from the nucleus,
pushing the inner disk out of 
alignment on a small, $\sim 100$ pc scale.

We now consider candidate scenarios which 
could account for such a warp:
\begin{enumerate}
\item Precession in the inner disk has
been induced by a binary massive black hole.
\item Radiation pressure
from the central source has twisted the disk.
The inner disk may have become warped via 
a self-induced radiation-driven mechanism 
\citep{pringle}.
\item There is strong coupling between the outer disk
and the inner disk, aligned by the Bardeen-Peterson
effect with the spin of the black hole \citep{natarajan98,scheuer}.
\item A jet-excited ambient interstellar medium 
has perturbed the disk. This would be a form of feedback from 
an AGN.
\end{enumerate}
We discuss these scenarios below. 

\subsection{Precession due to a binary black hole}

It is useful to summarize bulge and black hole properties
for 3C 449. \citet{bettoni03} have listed a bulge velocity
dispersion of 225 km~$s^{-1}$ and estimated 
a black hole mass of $2.5 \times 10^8 M_\odot$.
A transition radius (e.g., the sphere of influence)
where the gravitational force
from the nuclear black hole dominates that from the bulge is
\begin{equation}
\label{rt_def}
r_t =   {GM_{bh} \over 2 \sigma_*^2} =  10 {\rm pc }
          \left({M_{bh} \over 2.5 \times 10^8 M_\odot }\right)
          \left({225 {\rm km/s}  \over \sigma_*  }\right)^2
\end{equation}
%
Note that the transition radius can be compared to
the semi-major axis for a possible black hole binary which makes it ``hard''
\begin{equation}
a_{hard} < {G\mu \over 4 \sigma_*^2} = r_t {\mu \over 2 M_{bh}},
\end{equation}
where $\mu$ is the reduced mass of the black hole
binary and $M_{bh}$ now represents
the sum of the two black hole masses \citep{caproni}.
For an equal mass binary $a_{hard} = r_t/4 \sim $ a few pc 
for 3C 449.

We consider a disk exterior to a binary black hole 
in a circular orbit with semi-major axis
$a_b$ and reduced mass $\mu$. 
At radius $r>a$, 
the binary black hole exerts a quadrupolar perturbation on the 
gravitational potential
\begin{equation}
\Phi_2 =  {3 G \mu a_b^2 z^2 \over r^5}. 
\end{equation}
The vertical oscillation frequency 
is estimated by differentiating $\Phi_2$ twice with
respect to $z$, and adding it to that from 
the spherical component from the galaxy,
\begin{equation}
\nu \sim  \Omega \left[ 1 
             +  { 3 \mu \over M(r)}
                 \left({ a_b \over r }\right)^2 
           \right],
\end{equation}
where $M(r)$ is the total mass within radius $r$.
The precession rate is
\begin{equation}
\dot \delta = \nu - \Omega  \sim 
               \Omega 
                \left({ 3\mu \over M(r)}\right) 
                \left({ a_b \over r }\right)^2.
\end{equation}
Relating $M(r)$ to the velocity of a particle
in a circular orbit, we have $M(r) = v_c^2 r/G$.
Outside the black hole sphere of influence, we expect
$M(r) \propto r$, and $\Omega \propto r^{-1}$, 
consistent with a flat rotation curve, such that 
$\dot \alpha \propto r^{-4}$.
The precession period is then
\begin{eqnarray}
P_\delta & \sim &  10^{10} {\rm yr}
   \left({\mu \over 10^8 M_\odot }\right)^{-1}
   \left({r \over 100 {\rm pc} }\right)^{4} \nonumber \\
   \qquad && 
   \left({a_b \over 1 {\rm pc} }\right)^{-2}
   \left({v_c \over 320 {\rm km/s} }\right).
\end{eqnarray}
(For an isothermal sphere $v_c \sim \sqrt{2} \sigma_*$.)

Such a long timescale and steep dependence on radius 
presents a problem for this scenario.
For a binary black hole semi-major axis larger than a few pc, dynamical
friction can bring the black holes together on a timescale
of order a few times the rotation period.
Only if the black hole binary is hard ($a_b \lesssim$ a few pc) 
can the black hole separation shrink slowly, within a timescale on the 
order of a Hubble time (e.g., \citealt{milos}).
If the black hole binary is not hard, then 
it would spiral inward too fast to create the warp.
We find that the estimated precession
timescale is too long to account for the warp in the disk.

One might consider alternative models where the disk
precessed while a secondary black hole
spiraled inward.  Since the black hole orbit need not be planar,
it could account for a highly warped geometry in the disk. 
In this case, however, one would expect to see disturbed material
on both large and small scales.  Other than the active nucleus,
the galaxy appears dynamically relaxed, and there is little evidence
for a recent galactic merger in its isophotal structure (see Fig.~\ref{fig:nic}). 

\subsection{Radiative Warp Instability}

Here we explore the proposal by \citet{pringle,maloney96} that the absorption
of radiation from a central source can cause a surrounding disk
to precess and warp.  
The precession rate
can be estimated from the constant torque solution of \citet{maloney96} 
(their equation 8), giving a precession period of
\begin{equation}
P_{RW} = {  48 \pi^2 \Sigma r^2 v_c c \over L_{bol}}
\end{equation}
where $\Sigma$ is the disk mass surface density,
and $L_{bol}$ is the bolometric luminosity of the central source.
Inserting values appropriate for 3C 449,
we estimate a precession period
\begin{eqnarray}
P_{RW} &=& 9 \times 10^{10} {\rm years}
\left({v_c           \over 330 {\rm km/s}       }    \right)
\left({\Sigma        \over 1 M_\odot/{\rm pc}^2}     \right) \nonumber \\ 
\qquad && \left({L_{bol}  \over 10^{42}{\rm erg/s}}     \right)^{-1}
\left({r        \over 100{\rm pc}}     \right)^2.
\end{eqnarray}
The values used are based on the following estimates:
The radio core luminosity at 5 GHz  is lower than
that estimated from an unresolved source in the nucleus
at optical wavelengths, with a luminosity of $\sim 10^{40.8}$ erg/s 
\citep{chiaberge99,capetti02},
and at X-ray wavelengths with a luminosity of $\sim 10^{41}$ erg/s 
\citep{donato}.  The total bolometric luminosity from
the core is estimated to be 10 times that
in X-ray, or $L_{bol} \sim 10^{42}$ erg/s \citep{donato}.
Assuming constant {\it R}-band opacity for a face on 
disk of 0.065, we use the extinction coefficients in \citet{mathis90} and 
estimate the surface mass density in the disk to be 
$\Sigma \approx 1.3 M_{\odot}$~pc$^{-2}$.  
The circular velocity $v_c \sim 330$km/s is estimated from
the velocity dispersion listed  by \citet{bettoni03},  
assuming an isothermal sphere.

A quarter precession period would be long enough to account for the 
extent of the
warp in 3C 449, and this would reduce the above timescale by a factor of a few.
Even so, this reduced timescale is still very long, so this model is not likely 
responsible for the warp. 
The precession rate might have been faster if 
the luminosity were significantly higher in the past. However, 
there should also be a corresponding increase in the surface density 
of the disk, consistent
with the higher rate of accretion.  We conclude
that precession due to radiation pressure from a central
source is probably too slow to generate the warp.  The growth time
for the radiative warp instability is approximately the same
as the above period divided by the order of the mode grown. However, the observed
warp is large-scale (with wavevector similar to the radius), 
so a higher order mode is unlikely to account for it.
This finding is consistent with a similar estimate by \citet{ferrarese99}, 
for the warped disk in NGC 6251.

\subsection{Quick Disk/Black Hole Alignment }

\citet{scheuer,natarajan98} showed that a black hole could align
with a disk on a timescale set by the ratio of
the black hole and disk angular momentum times the 
Lense-Thirring precession period.  The Lense-Thirring precession
period is $\propto r^3$ and the disk angular momentum
is  $\propto r$, so the timescale for alignment becomes
very long at distant radii. The ratio of the black hole angular momentum
to the disk angular momentum is not small enough to reduce
the alignment timescale at 100 pc to a level that would allow
alignment to take place within a Hubble time (see the Lense-Thirring
precession rate given above in Equation \ref{eqn:LT}). We can therefore 
reject this scenario based on these simple timescale estimates.

\subsection{Disk-ISM interaction}

If the pressure gradient from the ambient ISM across the disk is large, 
it can overcome the torque from the galaxy \citep{quillen99,quillen97}.
The pressure required to keep a disk of surface mass
density $\Sigma$ from precessing in a galaxy of ellipticity
$\epsilon_\Phi$ is 
\begin{eqnarray}
P_\tau & \gtrsim &
2 \times 10^{-11} {\rm dynes \ cm}^{-2}
\left({\epsilon_\Phi \over 0.05}                     \right)
\left({\Sigma        \over 1 M_\odot/{\rm pc}^2}     \right) \nonumber \\
\qquad &&
\left({v_c           \over 330 {\rm km/s}       }    \right)^2
\left({100 {\rm pc} \over r}\right)
\cos{(\theta_g)} \sin{(\theta_g)}
\label{xraypressure}
\end{eqnarray}
where
$\theta_g$ is the disk inclination angle (with respect to the galaxy axis)
\citep{quillen99}. 

\citet{hardcastle98} observed that the X-ray emission in 3C 449 on arcminute
scales was anisotropic, with isophotes aligned orthogonally
to the jet axis, suggesting that there is a deficit 
of X-ray emission near the radio lobes.
The central gas density is $\sim 4.6 \times 10^{-3} {\rm cm}^{-3}$,  temperature
$kT= 1.2$ keV, and central pressure of $10^{-12} {\rm Pa}= 
10^{-11} $ dyne~cm$^{-2}$  
at a radius of about 0.5 arcminute.
Consequently the pressure at a radius of few arcseconds 
from the nucleus could be a factor
of 10 or so higher, or $ \sim 10^{-10}$ dyne~cm$^{-2}$.
A comparison between this number and the relation
of Eqn.~\ref{xraypressure} shows that the pressure in
the X-ray emitting ISM is large enough to perturb 
the gas disk on reasonable timescales, so long as 
the isobars of X-ray emitting gas in the ISM are very elongated, 
structured, and anisotropically distributed.

\citet{quillen99} explored a simplistic model in which the ambient ISM
has isobars elongated along the jet axis. In this case,
the ISM exerts a torque on the disk, and gives rise to 
precession about the jet axis. However, if our model is correct, the
disk in 3C 449 passes smoothly from the galaxy midplane at its outer edge
to the plane perpendicular to the jet.
If the jet causes a wind,  
the disk would be lifted out of the plane perpendicular to the jets, 
rather than pushed into it 
(e.g., \citealt{quillenwind}). 
Similarly, an azimuthal pressure
variation across the disk may overcome the influence of the galaxy potential and
perturb the disk at 100 pc, though not in ways capable of creating the warp
we predict exists in 3C 449. As discussed by \citet{quillen99},
gas isobars initially aligned with the jet should align with the galaxy
equipotentials on a sound crossing time, of order the rotation period at 
100 pc (a few Myr). Therefore the torque on the disk provided 
by static gas pressure in the ISM,
no matter how strong or non-axisymmetric, could only cause the disk to precess and 
eventually settle into the galactic midplane, and not the plane perpendicular to the jets.
The precessing disk may be damped via internal kinematics 
(e.g., high viscosity, collisions, etc.), but the time associated with the 
damping would be of order the accretion timescale, which is too long to 
account for the warp. 
Even if the region is ram pressure dominated and the gas in the disk 
is stirred near the sound speed, the azimuthal pressure gradient required 
to warp the disk would be unrealistically large. 
The \citet{quillen99} model is
therefore incapable of explaining the existence of the warp in 3C 449. 
That study, however, did not take into account the emergence of {\it Chandra} 
observations revealing temperature fluctuations and buoyant bubbles 
in the X-ray emitting ISM surrounding an 
AGN (e.g., \citealt{fabian03}). Such structure in the region 
is likely associated with the jet lobes, responsible for injecting large 
amounts of energy and momentum into the ambient medium.
A drag force between jet-excited bubbles and the disk would not give rise 
to an instability (as explored by \citealt{quillenwind}), and the warp would 
be damped. The inner disk could thus be pushed into the plane perpendicular to the jet.  
If the UV emission traces an excited cone of material above the disk and oriented
along the jet axis, it may reveal the medium responsible for the torque that 
changed the alignment of the inner disk.

\section{Summary and Discussion}

In this paper we have examined a near-infrared/visible colormap
of the radio galaxy 3C 449, finding a twist in the isochromes.
This twist is unexpected, as a planar disk should have isochromes
aligned with the disk major axis and exhibiting reflective symmetry about
the minor axis.
The disk edge at a radius of $\sim 600$ pc 
has a major axis nearly (within $10^\circ$) coincident with
the galactic isophotes (as seen at 1.6 $\mu$m), suggesting
that the outer disk has settled to the midplane of
the galaxy. The isophotes at 1.6 $\mu$m reveal 
no evidence for a recent merger
or a strongly triaxial bulge. Consequently, the twist in
the isochromes is most likely due to the disk itself.

By integrating light through an absorbtive thin disk model,
we find that we can account for the twist in the isochromes
with a warped geometry. The model predicts rotation along the
minor axis of the outer disk that
is consistent with the sign of mean velocities measured
from spectroscopic observations by \citet{noelstorr}.
In relation to the warp model,
the pattern of UV and H$\alpha$+[NII] emission \citep{allen02,martel00}, 
brighter to the north of the nucleus, 
could be emission from above the disk, possibly in 
a wide cone oriented approximately along the jet axes.  
The UV emission in front of the disk and to the north of the nucleus
would be easier to see than an opposing cone 
behind the disk and to the south. This suggests 
a possible explanation for the lack of UV emission south of
the nucleus or associated with the disk on the western side.

A warp in a disk with a sharp
outer edge is not necessarily unexpected.
For example, 
NGC 7626's contains a dust lane that ends abruptly, 
suggesting that it has an outer edge similar to that
seen in 3C 449.
The disk in NGC 7626 is clearly warped (see images by \citealt{kleijn99}), 
and could resemble that of 3C 449 if viewed at a lower inclination. 
The ionized inner disk of NGC 6251 is tilted with respect to
its outer dusty disk that also exhibits a sharp outer edge 
\citep{ferrarese99}.

Our model for the warped disk of 3C 449 predicts that it 
is nearly parallel to the jet axis at a radius of 600 pc,  
but twists and becomes nearly perpendicular to the jet
at $\sim 100$ pc. We can find examples
of other radio galaxies which might exhibit similar warps.
For example, the disk in NGC 7626
is nearly edge-on, and twists such that the disk
is nearly perpendicular to the jets near the nucleus (see
images from \citealt{kleijn99}).  
3C 465, NGC 5127 and NGC 7052 contain disks with sharp outer edges.
For these disks, as is true for that of 3C 449, the pattern absorption
is almost triangular,
suggesting that there is a twist in the isochromes.  
The jets in NGC 7052 are $45^\circ$ 
from its disk, and the inner disk could be nearly perpendicular
to the jet (see images by \citealt{kleijn99}). If the disks 
in 3C 465 and NGC 5127 are twisted,
their inner disks would be nearly perpendicular
to their jet axes (see images by \citealt{dekoff00} and by \citealt{kleijn99}).
3C 236 contains an inner disk which is nearly
perpendicular to its jets, whereas the outer disk is tilted about
$25^\circ$ degrees from its outer disk \citep{dekoff00}.  
M84 (3C 272.1) has a similar inner and outer disk structure \citep{quillen99},
and again the inner disk is more nearly perpendicular to the jet axis.
NGC 6251 presents a counter example, as the inner ionized disk is not
oriented along the jet axis \citep{ferrarese99}.
In the future, our procedure for modeling warped dusty disks could
be used to study the disk geometries of these other objects. 

The notion that our warp model for 3C 449 accounts for jet/disk orthogonality 
is somewhat mysterious. 
Since the rotation period is shorter at smaller radii,
an initially misaligned disk should first settle 
into the galaxy midplane at its innermost edge. However, the disk
of 3C 449 is settled into the symmetry plane at large radii but {\it not}
in its inner regions.  
Such a morphology suggests that there
is a force, aligned with or associated with the jets or active
nucleus, that has pushed the disk preferentially at smaller radii.

We have considered 
the possibility that a torque associated
with a nuclear binary black hole could affect the disk
orientation. We have also noted that 
radiation pressure from a central source could cause the
disk to precess, or that the radiative warp instability proposed by \citet{pringle}
is playing a role. We have shown that the timescales associated with these scenarios
are too long to account for the warp in the disk of 3C 449.
The pressure in the X-ray emitting
ISM, however, is likely to be large enough that this medium could perturb
the disk. A static model with elongated isobars
would only cause the disk to precess, not be pushed
into the plane orthogonal to the jet axis. 
Such precession might be 
damped by processes internal to the disk (e.g., high viscosity, collisions, etc.), 
but the associated 
settling time would be of order the accretion timescale, which is again too long.
Non-axisymmetric winds driven off the disk by evaporative, radiative, and MHD processes
(like those discussed by \citealt{meyer}) are likely to play a role in the 
morphology of the disk, but not to a degree large enough to account for the warp.    
An interaction between the disk and the jet-excited ambient
ISM (resulting in a drag force or inflow) could cause
the disk to be pushed into the plane perpendicular to
the jet, so long as there is sufficient turbulence and anisotropy in the X-ray 
emitting region.    
This suggests the intriguing possibility that 
studies of nuclear disk morphology can be used 
to probe past interactions between outflow-excited material 
and a comparitively cold gaseous disk. 
In the case of 3C 449, the UV emission (visible only north of the nucleus) 
might help trace the intricacies of such an interaction.

A close look at images of disks in radio galaxies
reveals not only warps but lopsidedness, sharp edges, and spiral features.
Such observations are somewhat counterintuitive.
Because of differential rotation, a lopsided disk should rapidly
(on a few times the rotation period) become circular, 
a sharp edge should become smooth on a diffusion timescale (roughly
the rotation period divided by the viscosity alpha parameter),  
and a non-self-gravitating disk should only exhibit transient spiral features.
As such, the morphology of such 
disks suggests that they may have been recently
perturbed, even truncated.
Notably nonactive elliptical galaxies can contain
disks which exhibit similar structures.
It is possible that gaseous disks in elliptical galaxies
are perturbed by present or past nuclear activity.  
Future high angular resolution studies could better
probe this possible connection.

Such a relationship between nuclear disk morphology 
and the ambient ISM can be considered a specialized manifestation
of ``feedback'' from an AGN.    
Accreting black holes release enormous amounts of energy to their surroundings
in various forms (through heating, radiation pressure, and 
induced kinetic motions, as discussed by \citealt{begelman, fabian03}). 
It is not unreasonable, then, to interpret structurally perturbed 
nuclear disks as consequences of current and past interactions
between an AGN and its surrounding medium.

\acknowledgements
We gratefully acknowledge helpful discussions with 
David Merritt, Eric Blackman, Stephen Thorndike, Ivan Minchev,
Kate Green, Russell Knox, and Amanda LaPage. We also 
thank George Privon for reduction of {\it VLA} data for 3C 449, 
and the anonymous referee whose valuable comments led to the 
improvement of this paper. 
This work was based on observations with the NASA/ESA
{\it Hubble Space Telescope}, obtained in collaboration with the Space Telescope 
Science Institute (STScI), operated by AURA for NASA. 
Support for this work was provided by NASA/STScI through grant HST-GO-10173.   
G.~R.~T. and A.~C.~Q. acknowledge support in part by
NSF awards AST-0406823, PHY-0242483, and NASA/STScI grant HST-GO-10173.09-A. 
This research has made extensive use of the NASA Astrophysics 
Data System (ADS) and the
NASA/IPAC Extragalactic Database (NED), operated by the 
Jet Propulsion Laboratory, California Institute of Technology, 
under contract with the National Aeronautics and Space Administration.


\begin{thebibliography}{}

\bibitem[Allen et al.(2002)]{allen02}
Allen, M. G., Sparks, W. B., Koekemoer, A., Martel, A. R., 
O'Dea, C. P., Baum, S. A., Chiaberge, M., Macchetto, F. D.,
\& Miley, George K. 2002, ApJS, 139, 411	

\bibitem[Armitage \& Natarajan(1999)]{armitage}
Armitage, P. J., Natarajan, P.  1999, ApJ, 525, 909


\bibitem[Bardeen \& Petterson(1975)]{bardeen75}
Bardeen, J.~M., \& Petterson, J.~A.~1975, ApJ, 193, L65

\bibitem[Begelman(2004)]{begelman}
Begelman, M. C. 2004,
Coevolution of Black Holes and Galaxies, from the Carnegie Observatories Centennial Symposia. Published by Cambridge University Press, as part of the Carnegie Observatories Astrophysics Series. Edited by L. C. Ho,  p. 375.

\bibitem[Bettoni et al.(2003)]{bettoni03}
Bettoni, D., Falomo, R., Fasano, G., \& Govoni, F.
~2003, A\& A, 399, 869

\bibitem[Capetti et al.(2002)]{capetti02}
Capetti, A., Celotti, A., Chiaberge, M., de Ruiter, H. R., 
Fanti, R., Morganti, R., \& Parma, P. 2002, A\&A, 383, 104	

\bibitem[Caproni \& Abraham(2004)]{caproni}
Caproni, A., \& Abraham, Z.~2004, ApJ, 602, 625   

\bibitem[Chiaberge et al.(1999)]{chiaberge99}
Chiaberge, M., Capetti, A., \& Celotti, A. 1999, A\&A, 349, 77	

\bibitem[Chiaberge et al.(2003)]{chiaberge03}
Chiaberge, M., Gilli, R., Capetti, A., \& Macchetto, F. D. 2003, ApJ, 597, 166	

%
\bibitem[Cornwell \& Perley(1984)]{cornwell84}
Cornwell, T., \& Perley, R., 1984, in Bridle A. H., Eilek J. A., eds, 
Physics of Energy Transport in Radio Galaxies, NRAO Workshop No. 9. NRAO, 
Green Bank, West Virginia, p. 39


\bibitem[de Koff et al.(1996)]{dekoff96}
de Koff, S., Baum, S. A., Sparks, W. B., Biretta, J., Golombek, D.,
Macchetto, F., McCarthy, P., \& Miley, G. K.\ 1996, ApJS, 107, 621

\bibitem[de Koff et al.(2000)]{dekoff00}
de Koff, S., Best, P., Baum, S. A., Sparks, W., Rottgering, H., Miley, G.,
Golombek, D., Macchetto, F., \& Martel, A.\ 2000, ApJS, 129,  33

\bibitem[Donato et al.(2004)]{donato}
Donato, D., Sambruna, R. M., \& Gliozzi, M. 2004, ApJ, 617, 915	


\bibitem[Fabian et al.(2003)]{fabian03}
Fabian, A.~C., Sanders, J.~S., Allen, S.~W., Crawford, C.~S., Iwasawa, K.,
Johnstone, R.~M., Schmidt, R.~W., \& Taylor, G.~B.\ 2003, MNRAS, 344, 43

\bibitem[Fanaroff \& Riley(1974)]{fanaroff74}
Fanaroff, B.~L. \& Riley, J.~M. 1974, MNRAS, 167, 31p


\bibitem[Ferrarese et al.(1996)]{ferrarese96}
Ferrarese, L., Ford, H. C., \& Jaffe, W.~1996, ApJ, 470, 444

\bibitem[Ferrarese \& Ford(1999)]{ferrarese99}
Ferrarese, L., \& Ford, H. C.~1999, ApJ, 515, 583


\bibitem[Ferrarese \& Merritt(2001)]{ferrarese}
Ferrarese, L., \& Merritt, D.~2000, ApJ, 539, L9


\bibitem[Feretti et al.(1999)]{feretti99}
Feretti, L., Perley, R., Giovannini, G. \& Andernach, H.~1999, A\&A, 341, 29

\bibitem[Gebhardt et al.(2000)]{gebhardt2000}
Gebhardt, K. et al.~2000, ApJ, 543, L5

\bibitem[Gebhardt et al.(2003)]{gebhardt2003}
Gebhardt, K. et al.~2003, ApJ, 583, 92

\bibitem[Goudfrooij \& de Jong(1995)]{goudfrooij95}
Goudfrooij, P., \& de Jong, T.~1995, A\&A, 298, 784

\bibitem[Gower \& Hutchings(1982)]{gower82}
Gower, A., \& Hutchings, J.~B.~1982, ApJ, 258, L63

\bibitem[Hardcastle, Worrall, \& Birkinshaw(1998)]{hardcastle98}
Hardcastle, M. J., Worrall, D. M., \& Birkinshaw, M.\ 1998, MNRAS, 296, 1098

\bibitem[Jaffe et al.(1996)]{jaffe}
Jaffe, W., Ford, H., Ferrarese, L., van den Bosch, F., \& 
O'Connell, R. W.~1996, ApJ, 460, 214

\bibitem[Kinney et al.(2000)]{kinney}
Kinney, A.L., Schmidt, H.R., Clarke, C. J., Pringle, J.E., Ulvestad, J.S., \&
Antonucci, R.R.J. 2000, ApJ, 537, 152

\bibitem[Kotanyi \& Ekers(1979)]{kotanyi79}
Kotanyi, C., \& Ekers, R.~1979, A\&A, 73, L1

\bibitem[Kumar \& Pringle(1985)]{kumar}
Kumar, S., \& Pringle, J. E. 1985, MNRAS, 213, 435

\bibitem[Lauer et al.(2005)]{lauer05}
Lauer, T. et al.~2005, AJ, 129, 2138

\bibitem[Lense \& Thirring(1918)]{lense}
Lense, J., \& Thirring, H.~1918, Phys.~Z., 19, 156

\bibitem[Liu(2004)]{liu04}
Liu, F. K. 2004, MNRAS, 347, 1357	

\bibitem[Lubow et al.(2002)]{lubow}
Lubow, S. H., Ogilvie, G. I., \& Pringle, J. E.
2002, MNRAS, 337, 706


\bibitem[Madrid et al.(2005)]{snap}
Madrid, J.P., Chiaberge, M., Floyd, D., 
Sparks, W.F., Macchetto, D., Miley, G.K., 
Axon, D., Capetti, A., O'Dea, C., Baum, S., 
Perlman, E.~S., \& Quillen, A.~C.~2005, in preparation

\bibitem[Mathis(1990)]{mathis90}
Mathis, J.~S. 1990, ARA\&A, 28, 37

\bibitem[Macchetto, McCarthy, \& Miley(1996)]{macchetto96}
Macchetto, F., McCarthy, P., \& Miley, G.~K.\ 1996, ApJS, 107, 621

\bibitem[Maloney, Begelman \& Pringle(1996)]{maloney96}
Maloney, P.~R., Begelman, M.~C., \& Pringle, J.~E.~1996, ApJ, 472, 582

\bibitem[Martel et al.(2000)]{martel00}
Martel, A., Turner, N., Sparks, W., Baum, S.~2000, ApJS, 130, 267

\bibitem[McCarthy et al.(1997)]{mccarthy97}
McCarthy, P. et al. 1997, ApJS, 112, 415

\bibitem[Merritt \& Ekers(2002)]{merritt}
Merritt, D., \& Ekers, R. D.
2002, Science, 297, 1310


\bibitem[Meyer \& Meyer-Hofmeister(1994)]{meyer}
Meyer, F., \& Meyer-Hofmeister, E.~1994, A\&A, 288, 175


\bibitem[Milosavljevic \& Merritt(2003)]{milos}
Milosavljevic, M., \& Merritt, D.~2003, ApJ, 596, 860	

\bibitem[M\"{o}llenhoff, Hummell, \& Bender(1992)]{mollenhoff92}
M\"{o}llenhoff, C., Hummel, E., \& Bender, R.~1992, A\&A, 255, 35


\bibitem[Natarajan \& Pringle(1998)]{natarajan98}
Natarajan, P., \& Pringle J. E.~1998, ApJ, 506, L97



\bibitem[Noel-Storr et al.(2003)]{noelstorr}
Noel-Storr, J., Baum, S.~A., Verdoes Kleijn, G., van der Marel, R.~P.,
O'Dea, C.~P., de Zeeuw, P.~T., \& Carollo, C.~M.~2003, ApJS, 148, 419 

\bibitem[Noel-Storr et al.(2005)]{noelstorr05}
Noel-Storr, J., Baum, S.~A., O'Dea, C.~P. et al.~2005, in preparation


\bibitem[O'Dea et al.(2001)]{odea01}
O'Dea, C.~P., Koekemoer, A.M., Baum, S.A., Sparks, W.B., 
Martel, A.R., Allen, M.G., Macchetto, F.D., \& Miley, G.K.	
2001, AJ, 121, 1915	

\bibitem[Peng et al.(2002)]{peng}
Peng, C.~Y., Ho, L.~C., Impey, C.~D., \& Rix, H.~2002, AJ, 124, 266

\bibitem[Pringle(1996)]{pringle}
Pringle, J.~E.~1996, MNRAS, 281, 357

\bibitem[Prugniel \& Simien(1997)]{prugniel}
Prugniel, P., \& Simien, F.~1997, A\&A, 321, 111


\bibitem[Quillen \& Bower(1997)]{quillen97}
Quillen, A.~C., \& Bower, G.~1997, (astro-ph/9709107)


\bibitem[Quillen \& Bower(1999)]{quillen99}
Quillen, A.~C., \& Bower, G.~1999, ApJ, 522, 718

\bibitem[Quillen et al.(1999)]{quillencone}
Quillen, A. C., Alonso-Herrero, A., Rieke, M. J., McDonald, C., 
Falcke, H., \& Rieke, G. H. 1999, ApJ, 525, 685


\bibitem[Quillen(2001)]{quillenwind}
  Quillen, A.~C.~2001, ApJ, 563, 313


\bibitem[Rees(1978)]{rees78}
Rees, M.~J. 1978, Nature, 275, 516

\bibitem[Rees et al.(1982)]{ree82}
Rees, M.~J., Begelman, M.~C., Blandford, R.~D., \& Phinney, E.~S.~1982,
Nature, 295, 17

\bibitem[Rees(1984)]{ree84}
Rees, M.~J.~1984, ARA\&A, 22, 271

\bibitem[Sansom et al.(1987)]{sansom87}
Sansom, A.~E. et al.~1987, MNRAS, 229, 15

\bibitem[Scheuer \& Feiler(1996)]{scheuer}
Scheuer, P. A. G., \& Feiler, R. 1996, MNRAS, 282, 291	

\bibitem[Schmitt et al.(2002)]{schmitt}
Schmitt, H. R., Pringle, J. E., Clarke, C. J., \& Kinney, A. L.
2002, ApJ, 575, 150

\bibitem[Solrzano-Iarrea et al.(2002)]{solrzano}
Solrzano-Iarrea, C., Tadhunter, C. N., \& Bland-Hawthorn, J.	
2002, MNRAS, 331, 673	

\bibitem[Sparks et al.(2000)]{sparks}
Sparks, W. B., Baum, S. A., Biretta, J., Macchetto, F. D., \& Martel, A. R.	
~2000, ApJ, 542, 667

\bibitem[Steiman--Cameron \& Durisen(1988)]{ste88}
Steiman--Cameron, T.~Y., \& Durisen, R.~H.~1988, ApJ, 325, 26

\bibitem[Tran et al.(2001)]{tran01}
Tran, H.~D., Tsvetanov, Z., Ford, H.~C., \& Davies, J.~2001, AJ, 121, 2928

\bibitem[Tremaine et al.(2002)]{tremaine2002}
Tremaine, S. et al.~2002, ApJ, 574, 740

\bibitem[van Dokkum \& Franx(1995)]{vandokkum95}
van Dokkum, P.~D., \& Franx, M.~1995, AJ, 110, 2027

\bibitem[Verdoes Kleijn et al.(1999)]{kleijn99}
Verdoes Kleijn, G. A., Baum, S. A., de Zeeuw, P. T., \& O'Dea, C. P.
~1999, AJ, 118, 2592	

\bibitem[Verdoes Kleijn \& de Zeeuw(2005)]{kleijn}
Verdoes Kleijn, G. A., \& de Zeeuw, P. T.~2005, A\&A, 435, 43	

\end{thebibliography}
\end{document}